# Rotational energy levels in the ground vibrational state of methane with kHz-level accuracy from comb-referenced double-resonance and Lamb-dip spectroscopies

Vinicius Silva de Oliveira[1], Isak Silander[1], Hiroyuki Sasada[2,3], Sho Okubo[2], Hajime Inaba[2], Kevin K. Lehmann[4], and Aleksandra Foltynowicz[1]

[1]*Department of Physics, Umeå University, 901 87 Umeå, Sweden*
[2]*National Metrology Institute of Japan, National Institute of Advanced Industrial Science and Technology, Tsukuba, 305-8563, Japan*
[3]*Department of Physics, Keio University, Yokohama 223-8522, Japan*
[4]*Departments of Chemistry & Physics, University of Virginia, Charlottesville, VA 22904, USA*
*e-mail addresses: aleksandra.foltynowicz@umu.se; kl6c@virginia.edu; sasada@phys.keio.ac.jp*

Methane is a key spherical-top molecule, yet restrictive selection rules for one-photon transitions have prevented determination of its ground state (GS) energies with state-of-the-art kHz-level accuracy. We report the GS rotational energy level differences with kHz-level accuracy from two frequency-comb-referenced Doppler-free methods: optical-optical double-resonance spectroscopy in the Λ-type configuration, and Lamb-dip spectroscopy of allowed and forbidden transitions. A Hamiltonian fit to the data yields GS term values with rotational numbers up to $J = 12$ with kHz level accuracy.

The spectroscopy of methane, $CH_4$, has long been of intense fundamental and practical interest. Methane is the lightest stable spherical-top molecule, and plays an important role in the spectroscopic characterization of environments ranging from terrestrial to exoplanetary atmospheres [1,2]. While measurements with kHz-level precision are available for rotational transitions [3-6] and rovibrational transitions in the fundamental [7-9] and overtone [10-13] C-H stretch bands, the energies of most ground-vibrational-state levels of methane remain comparatively poorly determined. For many levels, the standard errors exceed those of the most precise transition frequencies by more than an order of magnitude [14,15] due to restrictive selection rules that limit which levels can be connected by experimentally accessible transitions.

The rovibrational levels of $CH_4$ can be assigned the $T_d$ group symmetry labels $A_1$, $A_2$, E, $F_1$, and $F_2$, associated with meta ($A_{1,2}$), ortho ($F_{1,2}$), and para (E) nuclear spin functions with total nuclear spin angular momentum quantum numbers $\boldsymbol{I} = 2, 1, 0$, respectively [16]. The selection rules for electric dipole transitions are $A_1 \leftrightarrow A_2$, $E \leftrightarrow E$, and $F_1 \leftrightarrow F_2$, and transitions between levels with different $\boldsymbol{I}$ are forbidden. Traditional combination-difference (CD) analysis, based on pairs of one-photon transitions with a common upper level, allows determination of energy differences only between levels of the same symmetry and parity and with rotational quantum number $J$ differing by 2 or less. Thus, several low-lying rotational levels of the ground vibrational state, including the lowest level of each symmetry, cannot be connected to other ground vibrational state levels using one-photon spectroscopy CDs. This lack of connectivity ultimately limits the experimental determination of the term values of higher-lying levels.

This problem has been recently highlighted in a comprehensive study by Kefala *et al.* [14], who compiled experimental rovibrational and rotational transitions of methane from 96 literature sources published up to July 2023. Using the MARVEL (Measured Active Rotational Vibrational Energy Levels) [17] procedure, they derived an extensive list of empirical energy levels up to 10,000 $cm^{-1}$ with $J \leq 25$, including uncertainties. In MARVEL, combinations of accurate experimental transition frequencies are used to determine differences between energy levels. However, because the available experimental data do not link levels of different symmetry and parity, the resulting energy levels fall into several disconnected networks. To place all networks on a common energy scale, the authors introduced ten artificial transitions, called "Magic numbers," that are assumed term values of selected ground vibrational state levels relative to the lowest ground state term value (defined as zero), taken from the rotational Hamiltonian predictions of Amyay *et al.* [18] (see Appendix D in End Matter for details). A similar problem existed for the water molecule, until Tobias *et al.* [19] used spectroscopic-network assisted precision spectroscopy (SNAPS) to connect the ortho and para rovibrational levels of water using an extensive network of Doppler-free transitions.

In this work, we use frequency-comb-referenced Doppler-free optical-optical double-resonance (OODR) spectroscopy in the Λ-type configuration (also referred to as stimulated emission probing [20]) to connect for the first time several low-lying levels of the GS of methane. The Λ-type transitions–in combination with previously measured transition frequencies of the $\nu_3$ fundamental– connect GS levels with the same symmetry but opposite parity and $\Delta J$ up to 3, and thus connect several different networks and eliminate the need for five of the first six Magic numbers. Moreover, we use Lamb-dip spectroscopy of allowed and forbidden transitions to determine 60 new ground-state combination differences (GS-CDs) with $\Delta J$ = 0, 1, and 2. Finally, we fit a Hamiltonian model to all GS-CDs determined in this work, and previous infrared [7,8,11,21], microwave [3,5,6], and RF [4] data, and report GS term values up to $J$ = 12 with kHz-level accuracy.

FIG. 1 shows a schematic of the energy levels of the methane C-H stretch bands addressed by the two spectroscopic methods. FIG. 1a) shows how a combination of $\Lambda$-type OODR spectroscopy (GS $\rightarrow 2\nu_3 \rightarrow \nu_3$) and one-photon spectroscopy (GS $\rightarrow \nu_3$) allows connecting ground state levels of different parity ($C_1$ and $C_2$) and with a difference of rotational quantum numbers $\Delta J \leq 3$. In the $\Lambda$-type OODR experiment, a near-infrared (NIR) field excites the GS $\rightarrow 2\nu_3$ transition ($\nu_{ad}$) between rovibrational levels $a \leftrightarrow d$, and a mid-infrared (MIR) field dumps the $2\nu_3 \rightarrow \nu_3$ transition ($\nu_{dc}$) between rovibrational levels $d \leftrightarrow c$. Transition frequencies in the $\nu_3$ band ($\nu_{bc}$) [7,8] and the R branch of the $2\nu_3$ band ($\nu_{ad}$) [12,13] have previously been measured with accuracy of the order of or below 1 kHz using comb-referenced Lamb-dip spectroscopy. Here, we determine accurate frequencies of the $2\nu_3 \rightarrow \nu_3$ transitions, which–together with the previously measured transition frequencies–allow determination of the GS-CDs as $\delta\nu_{GS} = \nu_{ad} - \nu_{dc} - \nu_{bc}$. FIG. 1b) shows the GS-CDs between the allowed ($\nu_{ac}$) and forbidden ($\nu_{bc}$) $\nu_3$ transitions measured using the Lamb-dip spectroscopy and used to determine accurate values for $\delta\nu_{GS} = \nu_{ac} - \nu_{bc}$ with $\Delta J \leq 2$.

The $\Lambda$-type OODR experiments were performed using the spectrometer recently described in detail in Ref. [13], based on a NIR external cavity diode laser (ECDL, Sacher Lion P-1650) tunable around 1.65 µm, and the MIR idler of an optical parametric oscillator (OPO, TOPTICA TOPO) tunable around 3.3 µm. A methane gas sample at pressure of ~1 Pa is contained in a 60-cm-long optical cavity resonant for the NIR laser (99.7% mirror reflectivity) and transmitting the MIR idler (cavity transmission around 70%). The MIR and NIR beams are spatially overlapped before the cavity and separated after the cavity using dichroic mirrors; their Rayleigh range is equal to 46 cm.

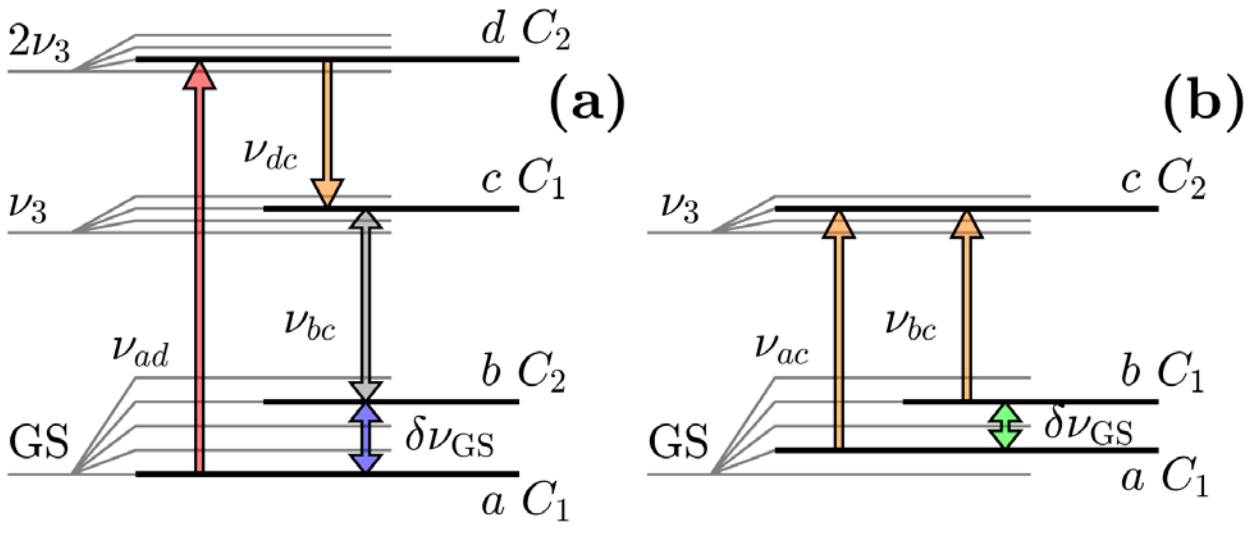

FIG. 1. (Color online) Schematics of methane rovibrational energy levels addressed in this work. a) In OODR, the NIR and MIR fields (red and orange arrows, $\nu_{ad}$ and $\nu_{dc}$) drive a $\Lambda$-type transition GS $\rightarrow 2\nu_3 \rightarrow \nu_3$; the grey arrow ($\nu_{bc}$) shows the previously measured $\nu_3$ transition, and the blue arrow shows the GS combination difference ($\delta\nu_{GS}$) determined here. b) In Lamb-dip spectroscopy, $\delta\nu_{GS}$ (green arrow) is obtained from a sequential measurement of allowed and forbidden $\nu_3$ transitions (orange arrows, $\nu_{ac}$ and $\nu_{bc}$). $C_{1,2}$ indicates the symmetry, $C$, and parity, 1 or 2, of the levels.

The NIR laser frequency ($\nu_{NIR}$) is locked to a $TEM_{00}$ mode of the cavity using the Pound-Drever-Hall (PDH) technique [22], and measured by referencing to a NIR frequency comb. The beat frequency between the NIR laser and the nearest comb line is kept constant by stabilizing the sample cavity length. The cavity resonance and the slave NIR laser frequencies are scanned across the $\nu_{ad}$ transition by changing the repetition rate of the NIR comb. The MIR idler is stabilized at a desired frequency by a phase-lock to a MIR frequency comb. The details of laser frequency stabilization are described in the Supplemental Material. The MIR beam intensity is modulated at 500 Hz using a mechanical chopper, and the intensity of the NIR beam transmitted through the cavity is demodulated at the modulation frequency. The linear absorption sensitivity of the NIR detection is $7.4\times10^{-10}$ $cm^{-1}\ Hz^{-1/2}$. Our previous work [13] demonstrated that this system could measure low-$J$ Lamb-dip $2\nu_3$ transition frequencies that agree with those reported by Votava *et al.* [12] to within 2 kHz. The NIR power transmitted through the cavity is ~100 µW, implying an intracavity power of ~30 mW. The MIR power incident on the cavity is in the range of 10 mW – 740 mW, depending on the transition.

During the $\Lambda$-type OODR measurements, the MIR laser frequency is locked at a detuning $\Delta\nu_{MIR}$ of a few MHz from the predicted $2\nu_3 \rightarrow \nu_3$ rest transition frequency, calculated as $\nu_{dc}^{\mathrm{pred}} = \nu_{ad} - \nu_{bc} - \delta\nu_{\mathrm{GS}}^{\mathrm{pred}}$, where $\nu_{ad}$ is taken from Refs. [12,13]; $\nu_{bc}$ from Refs. [7,8]; and $\delta\nu_{\mathrm{GS}}^{\mathrm{pred}}$ is calculated from the GS term values predicted by a previous spectroscopic fit of a rotational Hamiltonian [21]. Because of the MIR laser detuning from resonance, the $2\nu_3$ transition is split into two peaks, corresponding to the NIR beam co- and counter-propagating with respect to the MIR beam that have equal area but different widths [23], as explained in Appendix C in End Matter. The NIR laser is scanned in a ±15 MHz range around the rest frequency of the $2\nu_3$ transition in steps of 180 kHz; one scan lasts 1.6 min and we make multiple scans in opposite directions.

We measured 15 Λ-type OODR transitions, twelve allowing determination of $\delta\nu_{GS}$ with $\Delta J = 3$ for $J$ between 0 and 7, and three with $\Delta J = 1$ for $J$ between 1 and 4, which had significantly weaker dump transitions. FIG. 2 shows the Λ-type OODR spectrum of the $2\nu_3$ R(0, $A_1$) transition ($\nu_{ad}$ = 180345064527.50 ± 0.23 kHz [12]), observed when the MIR laser was locked at an offset $\Delta\nu_{MIR}$ = 4 MHz from predicted $2\nu_3 \leftarrow \nu_3$ P(2, $A_1$) transition frequency ($\nu_{dc}^{pred}$ = 88858195741 kHz). To determine the accurate frequency of the $2\nu_3 \leftarrow \nu_3$ transition, we fit a sum of two Lorentzian peaks to the OODR spectra, shown by the red curve. The residuals in the lower panel of FIG. 2 show that the lineshape model fits well except near the Lamb dip around $\Delta\nu_{NIR} = 0$, which is not included in the fit model. The Lamb dip has a derivative shape as, after lock-in demodulation, it appears as the difference of the Lamb dip with MIR radiation (which causes a small near resonant AC Stark shift) and the dip without MIR radiation. The fitting parameters are the center frequencies, widths and the total area of the Lorentzian peaks. The frequency splitting of the peaks is 16.791(34) MHz, and the half-width at half-maximum (HWHM) width is 1.823(62) MHz for the wider peak and 1.268(15) MHz for the narrower peak, where the main contribution to the width is the power broadening by the MIR laser [23]. The numbers in parentheses are the standard deviation of the results from fits to 10 consecutive scans. We determine the center frequency $\nu_{dc}$ of the $2\nu_3 \rightarrow \nu_3$ transition using Eq. (7) in End Matter, from a combination of the measured MIR idler frequency, the splitting of the NIR peaks, and $\nu_{ad}$ from Refs. [12,13]. The frequency is corrected for a second-order Doppler shift of ~ -150 Hz and recoil shift of ~ -1.1 kHz. We also correct for the power shift of the transition caused by the MIR laser, as described in the Supplemental Material. From multiple scans taken in opposite frequency directions and under different MIR detuning and power conditions, we determine $\nu_{dc}$ with accuracy of the order of a few kHz.

Finally, we calculate $\delta\nu_{GS}$ from the difference between $\nu_{ad}$ from Refs. [12,13], our experimentally determined $\nu_{dc}$, and $\nu_{bc}$ from Refs. [7,8]. For the example shown in FIG. 1a), the lower level $c$ of the $2\nu_3 \leftarrow \nu_3$ P(2, $A_1$) transition is the upper level of the $\nu_3$ P(3, $A_2$) transition ($\nu_{bc}$ = 89601828651.2 ± 2.5 kHz [7]). The CD between all three transitions allows to experimentally determine the separation of the ground state terms ($J$, $C$) = (0, $A_1$) and (3, $A_2$), which is also the Magic number 4. Appendix D in End Matter explains how other Magic numbers are replaced by our experimentally determined GS-CDs. Table S1 in the Supplemental Material lists the frequencies of all measured $2\nu_3 \rightarrow \nu_3$ transitions and the resulting $\delta\nu_{GS}$ with uncertainties of the order of a few kHz, calculated as combined uncertainties of the experimental $\nu_{dc}$ frequencies and the $\nu_{ad}$ and $\nu_{bc}$ frequencies from Refs. [7,8,12,13].

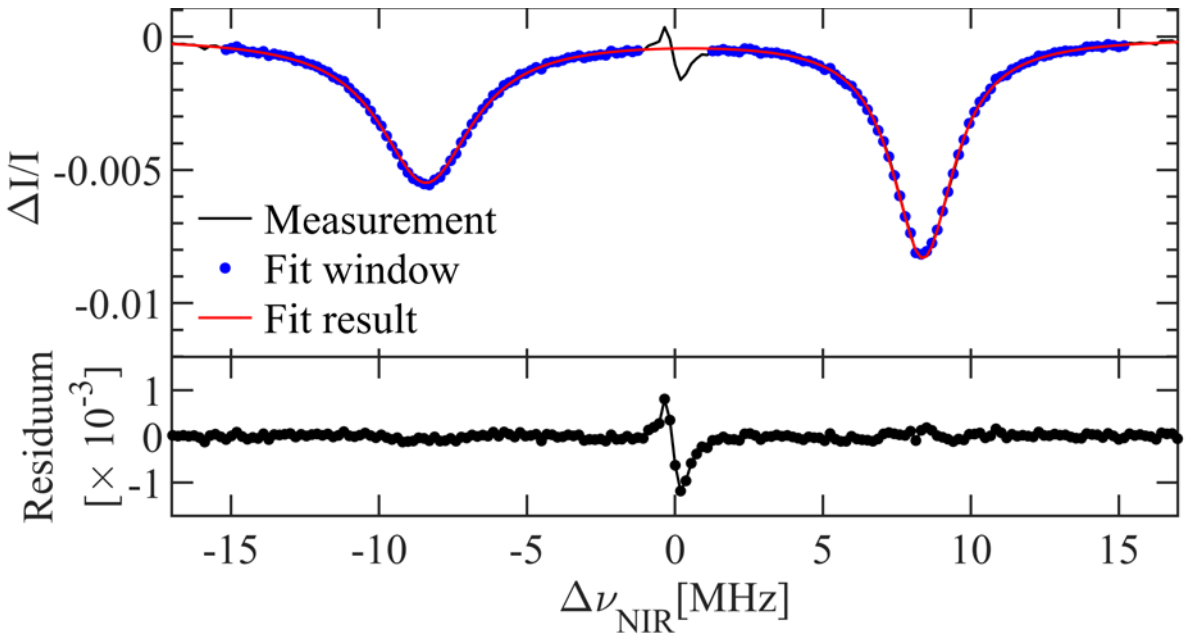

FIG. 2. (Color online) OODR spectrum of the $2\nu_3$ R(0, $A_1$) transition (single scan) measured with the MIR laser detuned by +4 MHz from the P(2, $A_1$) $2\nu_3 \leftarrow \nu_3$ transition. The narrower peak (higher frequency) corresponds to absorption of the NIR beam copropagating with the MIR beam. The markers show the experimental data, and the red curve shows the fit. The central Lamb-dip feature is not included in the fit. The bottom panel shows the fit residuals.

Precise Lamb-dip spectroscopy of the $\nu_3$ band was performed using an improved version of a spectrometer based on a cavity-enhanced absorption cell and a difference frequency generation (DFG) source [7,8]. The DFG source is now tunable up to 94.2 THz with output power level up to 30 mW. The 23.6-cm-long cavity (99.0% mirror reflectivity) contains 0.6 Pa of methane gas for the allowed transitions and from 0.6 Pa to 1.8 Pa for the forbidden transitions. The idler frequency is locked to the $TEM_{00}$ cavity mode using the PDH technique [22]. The power of the traveling wave in the cavity is from 10 to 50 μW for the allowed transitions and from 0.1 to 3 mW for the forbidden transitions. The Lamb dip of the allowed transitions is observed with a HWHM line width of about 0.3 MHz, which has contributions from transit time broadening (~100 kHz), pressure broadening (15-50 kHz, depending on the pressure), and power broadening, where the laser power was set to make the power broadening comparable to the combined transit-time and pressure broadening. A PZT attached to one of the cavity mirrors is modulated with a 5.2 kHz sinusoidal voltage and the transmitted wave is demodulated using a lock-in amplifier. FIG. 3 shows the $1f$ signal from the Lamb dip in a forbidden transition. The linear absorption sensitivity of the MIR detection is $5.4\times10^{-9}$ $cm^{-1}$ $Hz^{-1/2}$. For measurement of the transition frequencies, the cavity mode and the slave DFG idler frequency are locked to the center of the Lamb dip. The idler frequency is then measured by referencing the DFG pump and signal waves to a spectrally-broadened Er:fiber optical frequency comb. Details are described in the Supplemental Material.

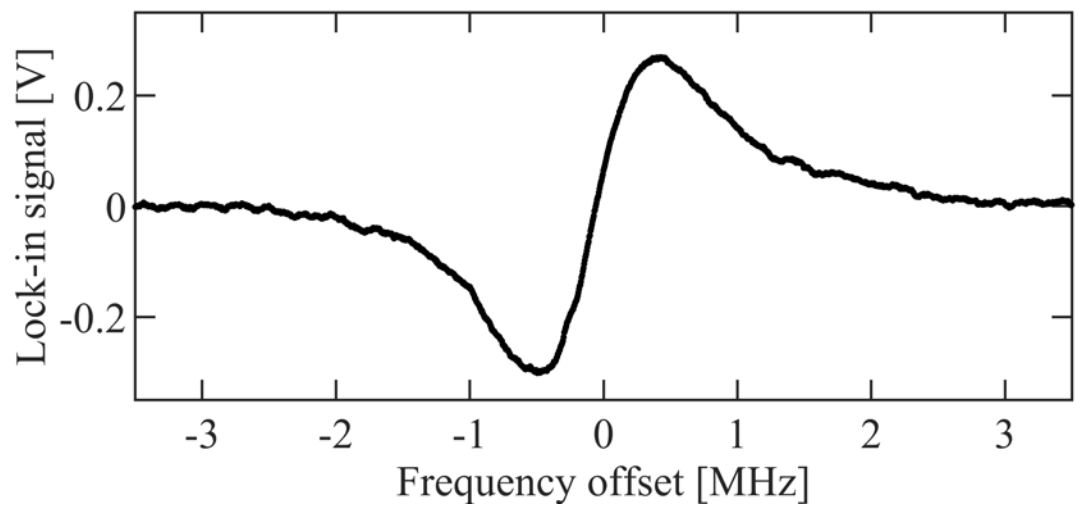

FIG. 3. (Color online) The 1*f* Lamb dip signal of the forbidden $\nu_3$ R(11, $F_2^{(3)}$) transition at 93.872 THz.

We determined Lamb-dip frequencies of 37 allowed transitions from R(9) to R(12) and 57 forbidden transitions from R(3) to R(12) in the $\nu_3$ fundamental band, with uncertainties ranging from 2 to 50 kHz, depending on the strength of the transitions. The uncertainties are calculated as the root sum square (RSS) of the standard deviations of the measured Lamb-dip frequencies, $\sigma_{meas}$, and the measured fluctuation of the beat frequency between the DFG signal and comb waves, $\sigma_{beat}$, as explained in the Supplemental Material. The value of $\sigma_{meas}$ dominates that of $\sigma_{beat}$ except for the very weak transitions.

Table S2 in the Supplemental Material lists all 84 GS-CDs with $J$ of 3 to 12 in the ground vibrational state, $\delta\nu_{GS}$, determined from the Lamb-dip frequencies of the forbidden and allowed transitions sharing a common upper level with the transitions measured in previous works [7,8,11,21]. The expected error of the GS-CDs is the RSS of the uncertainties of the forbidden and allowed transitions. Table S3 in the Supplemental Material lists 17 Lamb-dip frequencies of the allowed transitions that do not share the upper level with other observed Lamb-dip transitions. For consistency with the OODR data, we corrected all Lamb-dip frequencies, $\nu$, by the second-order Doppler shift of $-1.71 \times 10^{-12}$ $\nu$. Since the recoil shift causes a splitting of Lamb dips but no shift [24], no recoil correction was applied.

All GS-CDs determined from OODR and Lamb-dip spectroscopies are illustrated in FIG. 6 in End Matter. We used these data together with previous RF and microwave spectroscopy data in the ground and $\upsilon_3 = 1$ states [3-6,25], also marked in FIG. 6, as input to a weighted least squares fit of a model Hamiltonian in Eq. (1) in End Matter. The weight is inversely proportional to the square of the expected error. The number of the input data with $J \leq 12$ is 122: one from RF spectroscopy [4] and eight from microwave spectroscopy in the ground state [3,5], 16 from microwave spectroscopy in the $\upsilon_3 = 1$ state [6,25], 84 from Lamb-dip spectroscopy, and 13 from OODR spectroscopy (the last two lines in Table S1 are excluded from the OODR input data because the discrepancies between the measured and fitted frequencies are beyond three times the expected errors). Tables S1, S2, S4 and S5 in the Supplemental Material list all input data corrected by the recoil and second-order Doppler shifts and the residuals between the measured and fitted $\delta\nu_{GS}$. The residuals for the OODR and Lamb dip data are shown in FIG. 4 by the blue squares and green triangles, respectively, ordered by the lower state rotational quantum number up to $J = 6$. The error bars show the combined experimental and fit uncertainties. The chi-square of the weighted least squares Hamiltonian fit is 101.0, which is within a 90% confidence interval of 88.6 to 137.7 for the degree of freedom of 112. Table 1 in End Matter lists ten molecular constants determined from the fitted Hamiltonian. The numbers in the parentheses are the expected errors in the unit of the last digit, which are one order of magnitude lower compared to previous fit of the same Hamiltonian to a dataset that contained fewer Lamb dip GS-CDs and did not contain the OODR data [11].

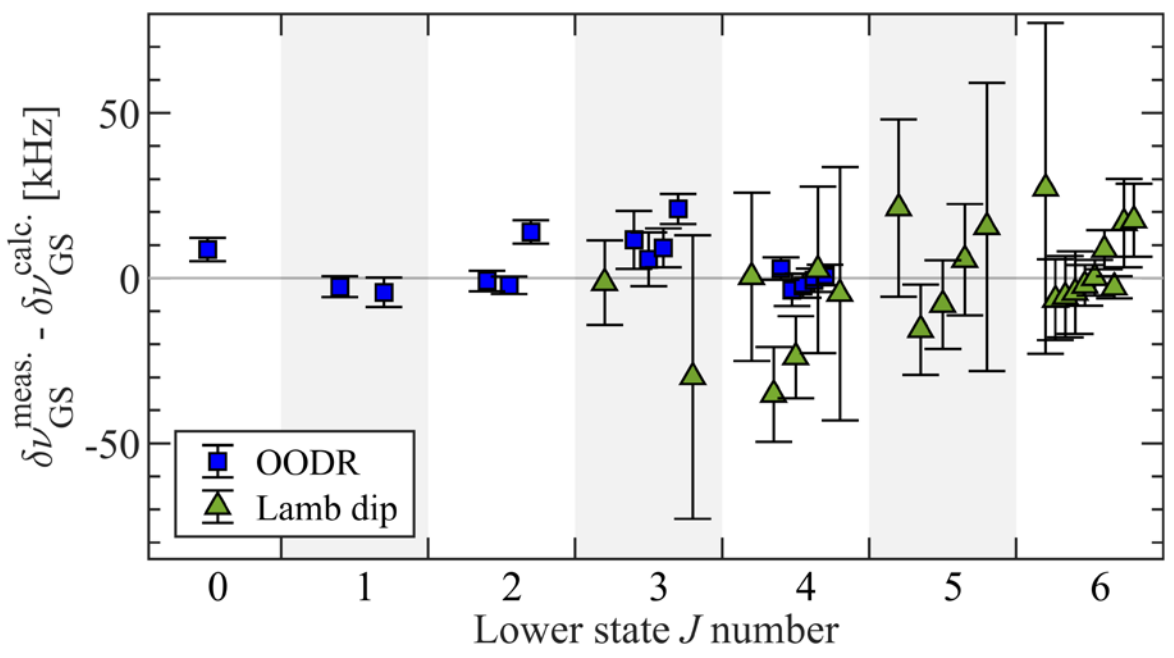

FIG. 4. (Color online) Frequency difference between the ground state combination differences, $\delta\nu_{GS}$, measured using OODR (blue squares) and Lamb dip (green triangles) spectroscopy and the predictions from the Hamiltonian fit.

Table S6 in the Supplemental Material lists the term values and the expected errors of the GS levels for $J \leq 12$ calculated using molecular constants from our Hamiltonian fit and the fit variance-covariance matrix. FIG. 5 shows the fitted GS energies compared to the values from Kefala *et al.* [14], retrieved from the ExoMol database [26], where the error bars are the combined uncertainties, dominated by the uncertainties from ExoMol (Fig. S2a in the Supplemental Material shows full error bars for levels with $J > 8$). The mean value of the difference is -458 kHz and standard deviation is 8.1 MHz. We believe the systematic discrepancies for levels with different symmetries are the consequence of the use of the Magic numbers, indicated by the black arrows, which fix some of the lowest levels to predicted rather than experimental values. Moreover, the uncertainties of these predicted levels, assumed to be 30 kHz in Ref. [14,26], are clearly underestimated, and more than an order of magnitude smaller than the experimental uncertainties of the $J > 4$ levels. The scatter of the $F_{1,2}$ and $A_{1,2}$ symmetry term values for $J > 4$ is low, which testifies to the accuracy of the experimental data used to link those levels to the Magic numbers. A similar comparison of our Hamiltonian GS energies to those from the HITRAN2024 [15] database yields randomly scattered residuals (see Fig. S2b in the Supplemental Material) with a mean of 474 kHz and standard deviation of 907 kHz, which is consistent with the rounding precision of $10^{-4}$ cm$^{-1}$ (3 MHz) in HITRAN.

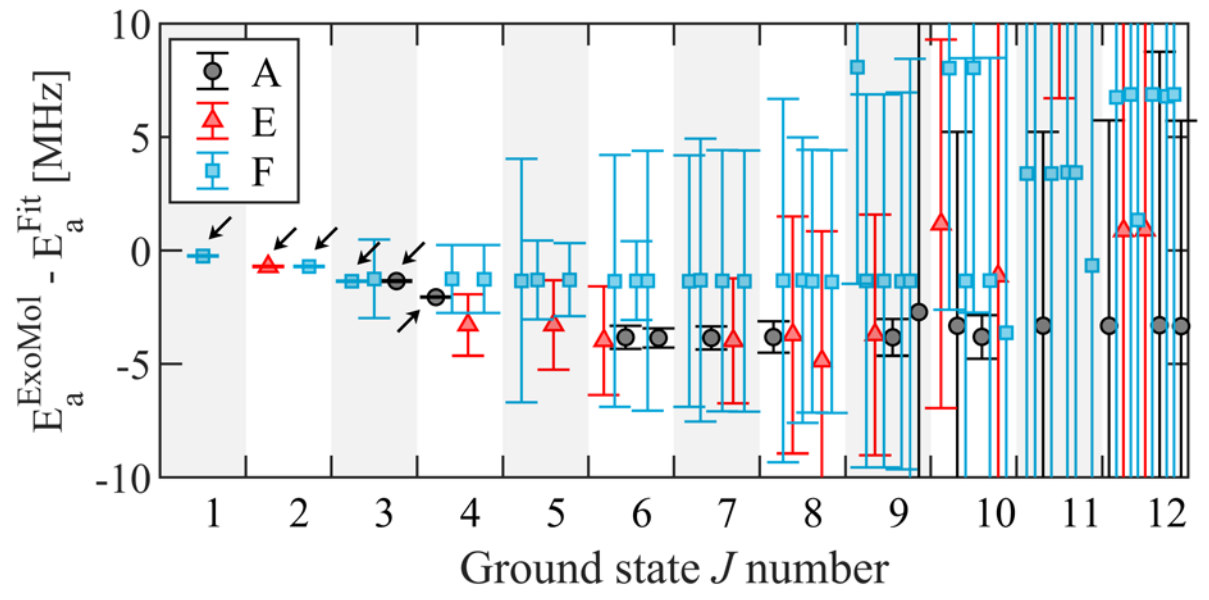

FIG. 5. (Color online) The difference between ground state energies from ExoMol and from the Hamiltonian fit, ordered by the rotational $J$ number and coded for the different symmetries as indicated in the legend. The levels indicated by black arrows are the Magic numbers.

In conclusion, we used two Doppler-free comb-referenced spectroscopic techniques to significantly expand the range of available experimental transitions linking the rotational levels of ground vibrational state of methane with kHz level accuracy. In particular, OODR spectroscopy in the Λ-type configuration allowed determination of CDs between levels with opposite parity and rotational quantum numbers $J$ differing by up to 3, while conventional one-photon CDs are limited to same parity levels and $\Delta J \leq 2$. Moreover, transitions involved in one-photon GS-CDs from low $J$ levels are too weak to be detected with Lamb-dip spectroscopy. Thus, the Λ-type OODR approach dramatically extends what levels can be connected by CDs. Our measurements, together with the previous RF data [4], eliminate the need for 5 out of 6 of the lowest Magic numbers and thus link 5 of the previously unconnected networks. The Hamiltonian fit to the combined data yields level energies with kHz level precision, orders of magnitude better than available in HITRAN and ExoMol. The standard errors of combined data are substantially reduced compared to previous fits without the OODR data. The GS term values from our work set a new benchmark for spectroscopic characterization of methane.

The authors thank Kyriaki Kefala and Jonathan Tennyson for useful discussions about the Magic numbers. A.F. acknowledges support from the Swedish Research Council (2020-00238) and the Knut and Alice Wallenberg Foundation (KAW 2020.0303). K.K.L. acknowledges funding from the U.S. National Science Foundation (CHE-2108458) and the Wenner Gren Foundation (GFOv2024-0010). H.S., S.O., and H.I. acknowledge funding from JST COI-NEXT (JPMJPF2015) and JSPS KAKENHI Grant Numbers 21H04500, 21H01852, and 25H00005.

[1] M. Hirtzig *et al.*, Icarus **226**, 470 (2013).

[2] M. R. Swain *et al.*, Nature **463**, 637 (2010).

[3] R. F. Curl, J. Mol. Spectrosc. **48**, 165 (1973).

[4] W. M. Itano and I. Ozier, J. Chem. Phys. **72**, 3700 (1980).

[5] M. Oldani, M. Andrist, A. Bauder, and A. G. Robiette, J. Mol. Spectrosc. **110**, 93 (1985).

[6] C. J. Pursell and D. P. Weliky, J. Mol. Spectrosc. **153**, 303 (1992).

[7] S. Okubo, H. Nakayama, K. Iwakuni, H. Inaba, and H. Sasada, Opt. Express **19**, 23878 (2011).

[8] M. Abe, K. Iwakuni, S. Okubo, and H. Sasada, J. Opt. Soc. Am. B **30**, 1027 (2013).

[9] P. A. Kocheril, C. R. Markus, A. M. Esposito, A. W. Schrader, T. S. Dieter, and B. J. McCall, J. Quant. Spectr. Rad. Transf. **215**, 9 (2018).

[10] C. Ishibashi, M. Kourogi, K. Imai, B. Widiyatmoko, A. Onae, and H. Sasada, Opt. Commun. **161**, 223 (1999).

[11] S. Okubo, H. Inaba, S. Okuda, and H. Sasada, Phys. Rev. A **103**, 022809 (2021).

[12] O. Votava, S. Kassi, A. Campargue, and D. Romanini, Phys. Chem. Chem. Phys. **24**, 4157 (2022).

[13] V. Silva de Oliveira, A. Hjältén, I. Silander, A. Rosina, M. Rey, K. K. Lehmann, and A. Foltynowicz, Opt. Express **33**, 38776 (2025).

[14] K. Kefala, V. Boudon, S. N. Yurchenko, and J. Tennyson, J. Quant. Spectr. Rad. Transf. **316**, 108897 (2024).

[15] I. E. Gordon *et al.*, J. Quant. Spectr. Rad. Transf. **353**, 109807 (2026).

[16] J. T. Hougen, in *International Review of Science, Physical Chemistry Series Two Volume 3 - Spectroscopy*, edited by A. D. Buckingham, and D. A. Ramsay (Butterworth, 1976).

[17] T. Furtenbacher, A. G. Császár, and J. Tennyson, J. Mol. Spectrosc. **245**, 115 (2007).

[18] B. Amyay, M. Louviot, O. Pirali, R. Georges, J. Vander Auwera, and V. Boudon, J. Chem. Phys. **144**, 024312 (2016).

[19] R. Tóbiás, T. Furtenbacher, I. Simkó, A. G. Császár, M. L. Diouf, F. M. J. Cozijn, J. M. A. Staa,

E. J. Salumbides, and W. Ubachs, Nat. Commun. **11**, 1708 (2020).
[20] M. Siltanen, M. Metsälä, M. Vainio, and L. Halonen, J. Chem. Phys. **139**, 054201 (2013).
[21] Y. Kobayashi, Master Thesis, Keio University, 2019.
[22] R. W. P. Drever, J. L. Hall, F. V. Kowalski, J. Hough, G. M. Ford, A. J. Munley, and H. Ward, Appl. Phys. B **31**, 97 (1983).
[23] K. K. Lehmann, J. Chem. Phys. **165**, 064118 (2026).
[24] J. L. Hall, C. J. Bordé, and K. Uehara, Phys. Rev. Lett. **37**, 1339 (1976).
[25] M. Takami, K. Uehara, and K. Shimoda, Japan. J. Appl. Phys. **12**, 924 (1973).
[26] J. Tennyson *et al.*, J. Quant. Spectr. Rad. Transf. **326**, 109083 (2024).
[27] K. T. Hecht, J. Mol. Spectrosc. **5**, 355 (1961).
[28] J. Moret-Bailly, L. Gautier, and J. Montagutelli, J. Mol. Spectrosc. **15**, 355 (1965).
[29] S. M. Kirschner and J. K. G. Watson, J. Mol. Spectrosc. **47**, 347 (1973).
[30] I. Ozier, J. Mol. Spectrosc. **53**, 336 (1974).
[31] C. L. Hu, J. Wang, T. P. Hua, A. W. Liu, Y. R. Sun, and S. M. Hu, Rev. Sci. Instrum. **92**, 073003, 073003 (2021).

### End matter

*Appendix A: Experimental GS term value differences* FIG. 6 shows the ground state rotational energy levels (not to scale) of methane up to rotational quantum number $J$ = 12. The blue and green arrows show the term separations determined in this work using OODR and Lamb-dip spectroscopy, respectively. The microwave and RF data used in the Hamiltonian fit are indicated by yellow, red and magenta arrows.

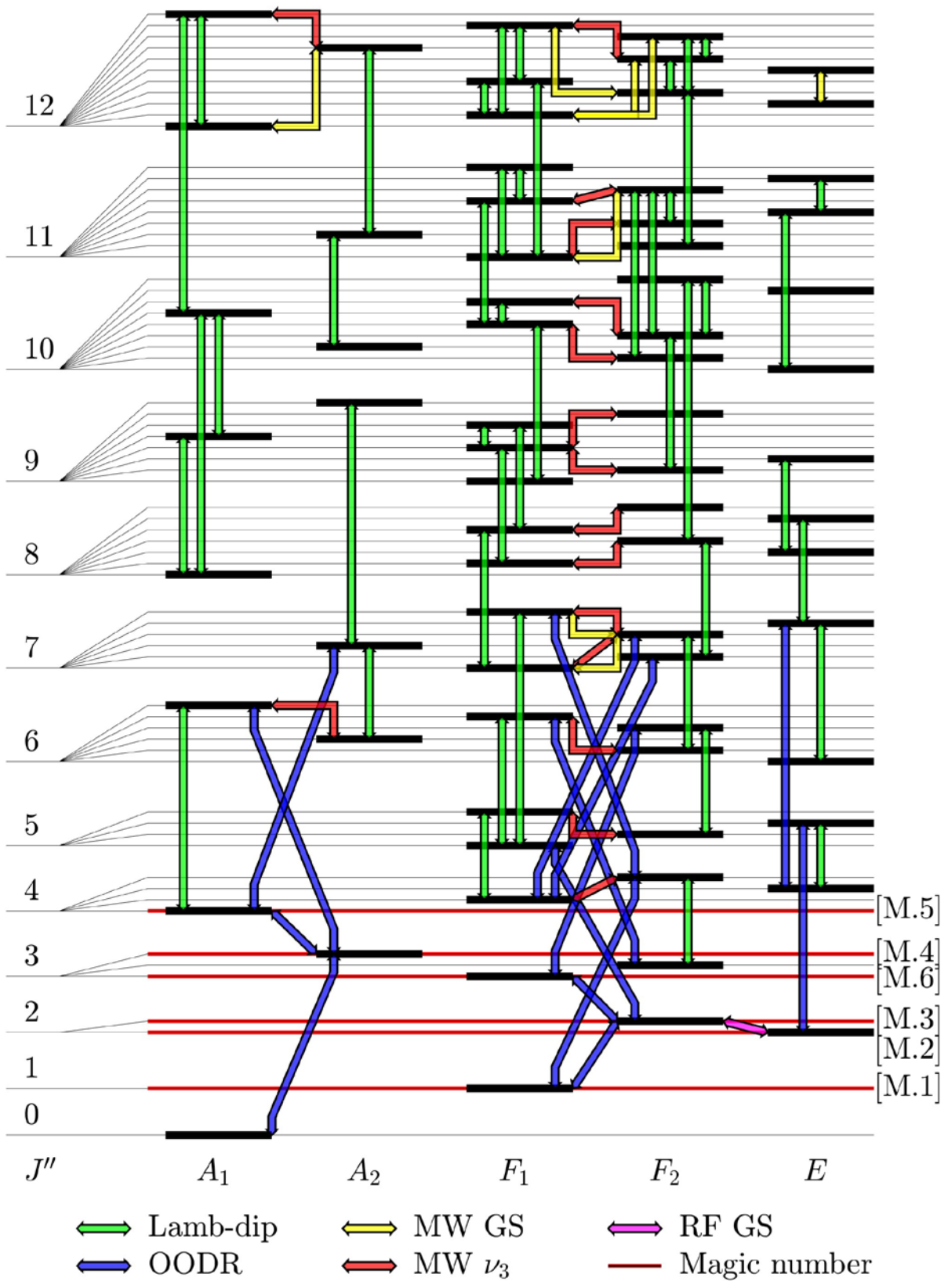

FIG. 6. (Color online) Schematic representation of the rotational levels of the ground vibrational state of methane up to $J$ = 12 and GS-CDs, $\delta\nu_{GS}$, determined using OODR via Λ-type transitions (blue) and using Lamb-dip spectroscopy of allowed and forbidden transitions (green). The term values that are Magic numbers are marked with red lines and M.1-M.6. The yellow and red arrows show CDs from the microwave data from Refs. [3,5,6,25] and the magenta arrow connecting the (2, $F_2$) and (2, E) levels shows the ortho-para splitting determined by RF spectroscopy in Ref. [4].

*Appendix B: Hamiltonian fit* The Hamiltonian used in the weighted least squares fit is given by

$$H/h = B\mathbf{J}^2 - D_s\left(\mathbf{J}^2\right)^2 + H_s\left(\mathbf{J}^2\right)^3 + L_s\left(\mathbf{J}^2\right)^4 + \left[D_t + H_{4t}\mathbf{J}^2 + L_{4t}\left(\mathbf{J}^2\right)^2\right]\Omega_4(J) + \left(H_{6t} + L_{6t}\mathbf{J}^2\right)\Omega_6(J) + L_{8t}\Omega_8(J) \quad (1)$$

where $h$ is the Planck constant, $\mathbf{J}$ is the non-dimensional angular momentum operator, and $\Omega_n$ is the $n$-th rank tensor operator [27-30]. The constants obtained from the fit are listed in Table 1 together with uncertainties.

TABLE 1. Molecular constants from the Hamiltonian fit.

| | Fit value [kHz] |
|---|---|
| $B$ | 157 122 620.40 (17) |
| $D_s$ | 3 328.679 1 (55) |
| $H_s$ / $10^{-1}$ | 1.921 40 (72) |
| $L_s$ / $10^{-5}$ | –1.705 (27) |
| $D_t$ | 132.943 72 (75) |
| $H_{4t}$ / $10^{-2}$ | –1.699 3 (13) |
| $L_{4t}$ / $10^{-6}$ | 2.066 (55) |
| $H_{6t}$ / $10^{-2}$ | 1.103 45 (59) |
| $L_{6t}$ / $10^{-6}$ | –2.707 (41) |
| $L_{8t}$ / $10^{-6}$ | –2.968 (90) |

*Appendix C: Lambda-type transition frequency* Consider the Λ-type OODR system shown in FIG. 1a), where the NIR field excites the $\nu_{ad}$ transition and the MIR field dumps the $\nu_{dc}$ transition. As was shown by Eqs. (5) and (8) in Ref. [31], the following relations hold for the NIR and MIR fields co- ($^+$) and counter-propagating ($^-$) with respect to each other

$$\frac{\nu_{ad}}{\nu^+_{\mathrm{NIR}}} - \frac{\nu_{dc}}{\nu_{\mathrm{MIR}}} - \alpha(\nu^+_{\mathrm{NIR}} - \nu_{\mathrm{MIR}}) = 0, \quad (2)$$

$$2 - \frac{\nu_{ad}}{\nu^-_{\mathrm{NIR}}} - \frac{\nu_{dc}}{\nu_{\mathrm{MIR}}} + 2\beta + \alpha(\nu^-_{\mathrm{NIR}} + \nu_{\mathrm{MIR}}) = 0, \quad (3)$$

where $\beta = \langle v^2 \rangle/(2c^2) = k_B T/(Mc^2) = 1.73\times 10^{-12}$ is the second order Doppler shift coefficient, $v = \sqrt{2k_B T/M}$ is the most probable velocity, $k_B$ the Boltzmann constant, $T$ the temperature, $M$ the rest mass of the molecule, and $c$ is the speed of light, $\alpha = h/(2Mc^2) = 1.387\times10^{-25}$ is the recoil shift coefficient, and $h$ is the Planck constant. For small detunings, i.e., $\Delta\nu_{\mathrm{MIR}} = \nu_{\mathrm{MIR}} - \nu_{dc} \ll \nu_{dc}$ and $\Delta\nu^{\pm}_{\mathrm{NIR}} = \nu^{\pm}_{\mathrm{NIR}} - \nu_{ad} \ll \nu_{ad}$, these equations can be series expanded as

$$\frac{\Delta\nu_{\mathrm{MIR}}}{\nu_{dc}} - \frac{\Delta\nu^{+}_{\mathrm{NIR}}}{\nu_{ad}} - \alpha(\nu_{ad} - \nu_{dc}) = 0, \qquad (4)$$

$$\frac{\Delta\nu_{\mathrm{MIR}}}{\nu_{dc}} + \frac{\Delta\nu^{-}_{\mathrm{NIR}}}{\nu_{ad}} + 2\beta + \alpha(\nu_{ad} + \nu_{dc}) = 0. \qquad (5)$$

Subtracting Eq. (5) from (4) and rearranging gives

$$\Delta\nu_{\mathrm{MIR}} = \frac{\nu_{dc}}{2\nu_{ad}}\left(\Delta\nu^{+}_{\mathrm{NIR}} - \Delta\nu^{-}_{\mathrm{NIR}}\right) - \beta\nu_{dc} - \alpha(\nu_{dc})^2, \qquad (6)$$

and finally

$$\nu_{dc} = \nu_{\mathrm{MIR}} - \frac{\nu_{\mathrm{MIR}}}{2\nu_{ad}}\left(\nu^{+}_{\mathrm{NIR}} - \nu^{-}_{\mathrm{NIR}}\right) + \beta\nu_{\mathrm{MIR}} + \alpha(\nu_{\mathrm{MIR}})^2, \qquad (7)$$

where $\nu_{dc}$ was replaced by $\nu_{\mathrm{MIR}}$ in the terms where the difference between these frequencies is negligible. When $\nu_{\mathrm{MIR}}$ is set slightly off resonance but within the Doppler width of the transition, the OODR signal observed while scanning $\nu_{\mathrm{NIR}}$ is split into two peaks separated by $\nu^{+}_{\mathrm{NIR}} - \nu^{-}_{\mathrm{NIR}}$. The transition frequency $\nu_{dc}$ is then calculated from this splitting and the MIR laser frequency, and corrected by the second-order Doppler shift and the recoil shift.

When the homogeneous widths of the transitions are considered, the $\nu^{+}_{\mathrm{NIR}}$ resonance will be taller and narrower than the $\nu^{-}_{\mathrm{NIR}}$ resonance with a ratio of approximately 3:5 [13,23] for $\nu_{\mathrm{NIR}} \approx 2\nu_{\mathrm{MIR}}$. Thus, the sign of $\nu^{+}_{\mathrm{NIR}} - \nu^{-}_{\mathrm{NIR}}$ is unambiguous from the experiment. This difference in width arises from the difference in Doppler shift of the two photon $a \rightarrow c$ transition depending on the direction of $k$ vectors combined with the finite width of the velocity group pumped due to the power broadening of the $d \rightarrow c$ transition.

*Appendix D: Magic numbers* The energy difference between the term values marked by M.1-M.6 in FIG. 6 and the $(J, C) = (0, A_1)$ level (defined as zero) are the six lowest Magic numbers introduced by Kefala *et al.* [14], taken from the predictions of a rotational Hamiltonian by B. Amyay *et al.* [18]. The ground $J$ = 0 level has $A_1$ symmetry. Thus, it is unsurprising that the term values for levels (1, $F_1$) and (2, E) are assigned as Magic.1 and Magic.2. Measurements made by Ozier *et al.* [4] (indicated by the pink arrow in FIG. 6) determined the splitting of the (2, $F_2$) and (2, E) levels, by observation of the anti-crossing of levels with different ***I*** in an external magnetic field. This measurement connects networks of E symmetry with those with $F_1$ and $F_2$ symmetries, and thus eliminates the need for Magic.2. The term values for (3, $A_2$) and (3, $F_1$) are labeled Magic.4 and Magic.6. The term value for (3, $F_2$) is not a Magic number, because GS-CDs between allowed and forbidden transitions allow the determination of the term value spacing between the (2, $F_2$) and (3, $F_2$) levels. For ground vibrational states with $J$ = 4, the spacings (2, E) – (4, E), (2, $F_2$) – (4, $F_2$) and (3, $F_1$) – (4, $F_1$) can be determined by one-photon CDs, but (4, $A_1$) cannot be connected to a lower level of $A_1$ symmetry; the only such level is (0, $A_1$); the term value of the (4, $A_1$) level is thus assumed as Magic.5. Experimental data exist to tie all the higher ground state rotational levels to those assigned the six Magic numbers until (20, E), which is Magic.7. The remaining four Magic numbers assign term values to several $J$ = 20 – 24 levels, which have very low thermal population at ambient temperature. The lowest of these, $J$ = (20, E), with an assumed term value of 2179.816 cm$^{-1}$, has a population of 7.13 $\times10^{-4}$ relative to the (0, $A_1$) level at ambient temperature.

Our measurements remove the need for 5 out of the six lowest Magic numbers. The term value for the (1, $F_1$) level, Magic.1, is connected to (4, $F_2$) via our OODR measurement series 2 (see Table S1 in the Supplemental Material), and to (2, $F_2$) via series 13. Series 14 determined the (2, $F_2$) – (3, $F_1$) spacing and thus eliminates the need for Magic.3. The term value for the (2, E) level, Magic.2, is connected via measurements of the splitting between (2, $F_2$) and (2, E) stated to be 7.97030(66) MHz [4]. The term value of the (3, $A_2$) level is directly connected to level (0, $A_1$) by OODR measurement series 1 and thus eliminates the need for Magic.4. The term value for the (4, $A_1$) level, Magic.5, is connected by our series 15 to (3, $A_2$); combined with transition series 1, this determines the (4, $A_1$) term value and thus eliminates Magic.5. The term value for the (3, $F_1$) level, Magic.6, is connected by series 14 to the (2, $F_2$). None of these Magic numbers could be resolved using conventional one-photon combination differences. In principle, direct rotational transitions between levels (1, $F_1$) – (2, $F_2$) (20.960725 cm$^{-1}$) and (3, $A_2$) – (4, $A_1$) (41.894652 cm$^{-1}$) could have resolved two of the Magic numbers but these are extremely weak transitions, with Einstein A coefficients of 1.88 × 10$^{-12}$ s$^{-1}$ and 1.75 × 10$^{-9}$ s$^{-1}$, respectively.

Thus, for the $J$ < 20 levels, the only remaining Magic number is the term value of the (1, $F_1$) level, which can be resolved only by measurement of a spacing between an $A_1$ or $A_2$ level relative to an E, $F_1$, or $F_2$ level. This could, in principle, be done by a Zeeman or Stark anti-crossing experiment, most promisingly between the levels (4, $A_1$) and (4, $F_1$) which are predicted by our fit to be separated by 55.7666(28) MHz, but this would require application of a highly homogeneous field of ~1.4 T.

## *Supplemental Material for*

# Rotational energy levels in the ground vibrational state of methane with kHz-level accuracy from comb-referenced double-resonance and Lamb-dip spectroscopies

Vinicius Silva de Oliveira[1], Isak Silander[1], Hiroyuki Sasada[2,3], Sho Okubo[2], Hajime Inaba[2], Kevin K. Lehmann[4], and Aleksandra Foltynowicz[1]

*[1]Department of Physics, Umeå University, 901 87 Umeå, Sweden*
*[2]National Metrology Institute of Japan, National Institute of Advanced Industrial Science and Technology, Tsukuba, 305-8563, Japan*
*[3]Department of Physics, Keio University, Yokohama 223-8522, Japan*
*[4]Departments of Chemistry & Physics, University of Virginia, Charlottesville, VA 22904, USA*
*e-mail addresses: aleksandra.foltynowicz@umu.se; kl6c@virginia.edu; sasada@phys.keio.ac.jp*

## I. EXPERIMENTAL DETAILS

### 1. OODR spectroscopy

**Laser frequency stabilization**

The NIR laser frequency is locked to one of the cavity resonances using the Pound-Drever-Hall technique [1]. The absolute frequency of the NIR laser, $\nu_{NIR}$, is measured by beating it with the shifted spectrum of a self-referenced 250 MHz Er:fiber frequency comb (Menlo Systems, FC-1500-250-WG), and the beat frequency is kept constant by feedback to the PZT that controls the sample cavity length. The cavity resonance and the slave NIR laser frequencies are scanned by changing the repetition rate, $f_{rep}$, of the NIR comb. The comb $f_{rep}$, $f_{ceo}$ and the beat frequency between the NIR comb and the NIR laser are monitored using counters referenced to a GPS-disciplined 10 MHz Rb oscillator (Symmetricom 4410A, relative uncertainty 6 $\times 10^{-12}$ in 1 s).

The idler of the MIR optical parametric oscillator (OPO) is phase-locked to a home-built 125 MHz $f_{ceo}$-free MIR frequency comb [2] by feedback to the current control of the OPO seed laser (CTL PRO, TOPTICA). The beat frequency between the OPO idler and the MIR comb is recorded during the measurements. A home-made FTS monitors both pump and signal waves of the OPO and the difference between these two frequencies determines which MIR comb tooth the idler is locked to.

**OODR signal demodulation**

The intensity of the transmitted NIR beam is detected using an amplified InGaAs photodiode (Thorlabs PDA20CS-EC) and demodulated at 500 Hz, the amplitude modulation frequency of the MIR beam. We record the in-phase and in-quadrature outputs of the lock-in amplifier (Stanford Research Systems, SR830) for post processing. The sign of the measured NIR signals indicates that the MIR laser causes an increase in the NIR laser absorption. The OODR signal is generated by the MIR laser depleting the steady state population of the upper level of the NIR transition and thus is proportional to the saturation parameter of the NIR transition, which in turn is predicted to scale inversely with the square of the sample pressure. We found that the Λ-type OODR signals could only be detected at low pressures (~2 Pa or less), unlike the case of ladder- or V-type OODR transitions observed in our previous work that used the MIR radiation as a pump and the NIR as a probe [3]. At these low pressures, we observe Lamb dips in the NIR transmission of a few percent fractional depths with half-width-at-half-maximum width of about 350 kHz.

Using the intensities for the $2\nu_3 \leftarrow \nu_3$ transitions from the ExoMol database [4,5], we found that the strongest such transitions were when state $b$ has $J$ three quanta higher than that of state $a$; these had calculated Einstein A coefficients larger than even the allowed $\nu_3$ fundamental lines.

**Pressure and power shifts**

Several systematic effects need to be considered that influence the transition frequencies. Equation (7) in the manuscript includes corrections for the second-order Doppler shift and the recoil shift, which are of the order of 150 Hz and 1.1 kHz, respectively.

To evaluate the pressure shift, we measured the $2\nu_3 \leftarrow \nu_3$ P(3, $F_1$) transition frequency at different sample pressures, as shown Fig. S1a), where the measured frequency ($\nu_{dc}^{\text{meas}}$) is offset by the predicted frequency ($\nu_{dc}^{\text{pred}}$). The slope of the line fitted to the data is (-0.6 ±2.7) kHz within 95% confidence interval, therefore no significant pressure dependence of the center frequency was observed within the uncertainty of our measurements. The pressure shift is expected to be dominated by the long-range dispersion interaction of the colliding molecules; as a $CH_4$ molecule pair has only an isotropic long-range interaction, the rotationally averaged result should be independent of rotational states. Moreover, Bagaev et al. [6] reported a Lamb-dip pressure shift of the P(7) $\nu_3$ line of $CH_4$ in He and Xe, two extremes of atomic collision partners, of below 400 Hz at 1 Pa. Therefore, we neglect the pressure shift contribution to the center frequency.

We observed, however, that the center frequency depended on the MIR laser power and the sign of the detuning from the rest transition frequency. To compensate for the power-induced frequency shift, each transition was measured at different MIR power levels and with both positive and negative MIR laser detuning. An example is shown in Fig. S1b) for the the $2\nu_3 \leftarrow \nu_3$ P(2, $A_1$) transition, where the x axis is expressed as the product of the transmitted power and the sign of the MIR laser detuning, $P_{\text{MIR}}^{\text{trans}} \Delta\nu_{\text{MIR}} / |\Delta\nu_{\text{MIR}}|$. The black solid line shows a linear fit to the data, and the black dashed lines represent the expanded uncertainty ($k = 2$) associated with the fit, $\sigma_{\text{fit}}$. The dashed red lines correspond to the 95% observational bounds, defined as the quadrature sum of the fit uncertainty and the spread of the individual measurement points, $\sigma_{\text{meas}}$,

$$\sigma_{\text{obs}} = \sqrt{\sigma_{\text{fit}}^2 + \sigma_{\text{meas}}^2} \tag{S.1}$$

For normally distributed noise, 95 % of the data points are expected to lie within the observational bounds, while the fitting uncertainty characterizes the uncertainty of the dataset as a whole. The center frequencies of the $2\nu_3 \rightarrow \nu_3$ transitions listed in Table S 1 are evaluated at zero power, and the reported uncertainty of the transition ($\sigma_{dc}$) is the fit uncertainty ($\sigma_{\text{fit}}$). To estimate the uncertainty of the ground-state combination difference ($\sigma_{\text{GS}}$), we combine the uncertainties of all three involved transitions in quadrature as

$$\sigma_{\text{GS}} = \sqrt{\sigma_{dc}^2 + \sigma_{ad}^2 + \sigma_{\text{bc}}^2} \tag{S.2}$$

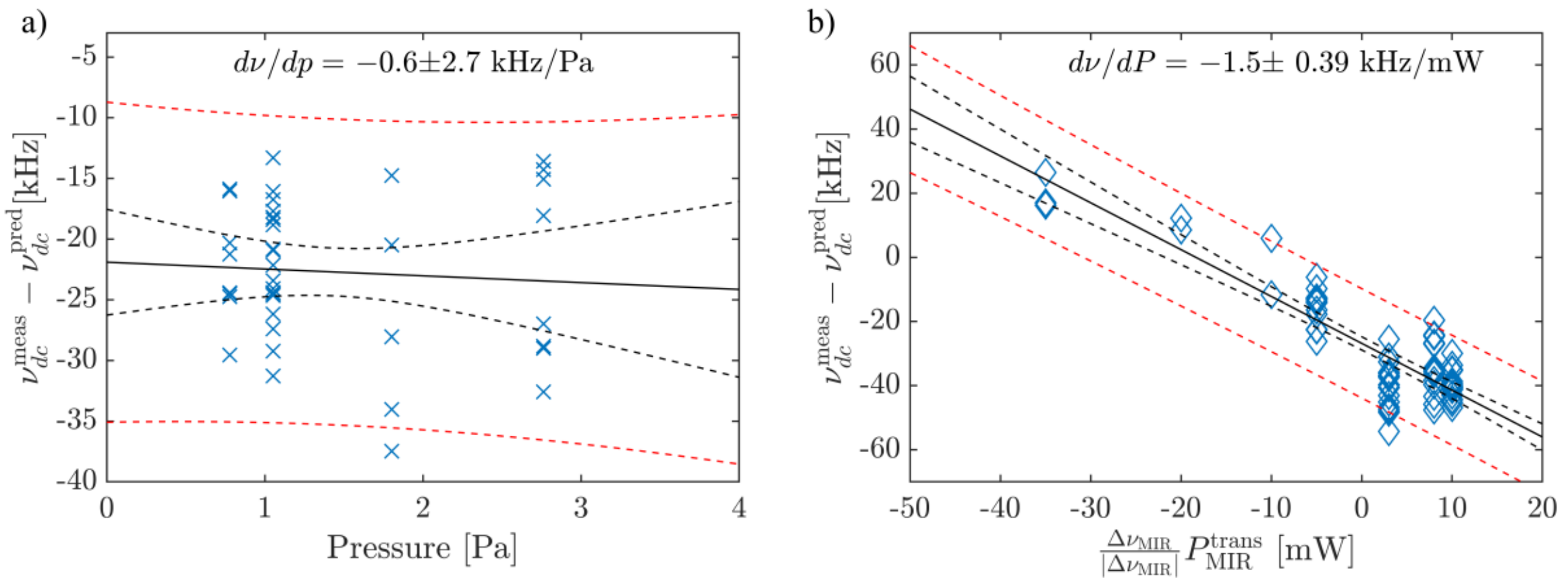

Fig. S1 (Color online) Difference between the measured ($\nu_{dc}^{\text{meas}}$) and predicted ($\nu_{dc}^{\text{pred}}$) transition frequency $\nu_{dc}$ as a function of (a) sample pressure for the $2\nu_3 \rightarrow \nu_3$ P(3, $F_1$) transition and (b) the MIR beam power and sign of detuning for the $2\nu_3 \rightarrow \nu_3$ P(2, $A_1$) transition. The solid black lines are the linear regression fits. The dotted black lines are the 95% confidence limits of the regression function. The dotted red lines are the 95% prediction limit, which include both the uncertainty of the linear regression and the scatter of the measurements.

**OODR measurement conditions**

To retrieve the center frequencies of the $2\nu_3 \rightarrow \nu_3$ transitions, we made multiple measurements of 15 Λ-type transitions. Each transition was measured with both positive and negative detuning $\Delta\nu_{\text{MIR}}$ of about 4 MHz, to allow correcting for the power shift. The incident MIR power was of the order of 10 to 140 mW for most transitions, but reached up to 740 mW for the weakest transitions. The pressure ranged from 0.64 to 1.38 Pa, depending on the strength of the transition. The methane sample flow through the optical cavity was controlled by a needle valve and a turbomolecular pump on opposite ends of the cavity. During measurements, we monitored the pressure in the cavity using a vacuum gauge (Leybold PTR90), and the MIR power transmitted by the cavity. For each transition, we recorded 23 to 110 scans at different combinations of MIR power and detuning. We fitted each scan individually and extracted the linear dependence of the frequency on the MIR power and detuning sign as illustrated in Fig. S1b).

## 2. Lamb-dip spectroscopy

### Laser frequency stabilization

The pump and signal sources of the difference-frequency-generation (DFG) source are a 1.06 μm external cavity diode laser (ECDL) and a 1.5 μm ECDL, respectively. The pump and signal waves are boosted with an Yb:fiber amplifier to 13 mW-400 mW and an L-band fiber amplifier to 40 mW-320 mW, respectively, and then led into a waveguide-type PPLN module with a conversion efficiency of 40%/W. The generated idler wave has a power level up to 30 mW and covers the frequency range from 91.7 THz to 94.2 THz.

The transition frequency of the Lamb dip is measured using a spectrally broadened Er:fiber optical frequency comb. The comb $f_{\mathrm{rep}}$ and the beat frequency between the DFG pump and comb waves are approximately 43 MHz and 10 MHz, respectively, and are phase-locked to individual RF synthesizers. These values are recorded with individual frequency counters. Simultaneously, the beat frequency between the DFG signal and comb waves is measured using two frequency counters to monitor count slips. The time base of the RF synthesizers and the frequency counters is UTC (NMIJ) with a relative uncertainty of $10^{-13}$ at a 1 s averaging time. The measurement gate time is 1 s, and the data are acquired and averaged at least 300 times per measurement set. We measured 3 to 12 sets for a particular transition, each set acquired during one day. The difference in the mode order of the comb is determined with a wavelength meter (Bristol Instrument 671). The signs of the beat notes are determined by slightly changing the repetition rate and the offset frequency between the pump and comb waves.

### Transition frequency uncertainty

We measured the Lamb-dip frequency, $\nu_i$ ($i$ = 1, 2, ...., $n$), of the transition $n$ times with the standard deviation of the beat frequencies between the signal and comb waves, $\delta f_i$. The frequencies in Table S 2 are the average of $\nu_i$, $\bar{\nu}$, and the uncertainties are given as

$$\sigma = \sqrt{\sigma_{\mathrm{meas}}^2 + \sigma_{\mathrm{beat}}^2} \tag{S.3}$$

where $\sigma_{\mathrm{meas}}$ is the standard deviation of $\nu_i$ given as

$$\sigma_{\mathrm{meas}}^2 = \frac{1}{n-1}\sum_{i=1}^{n}(\nu_i - \bar{\nu})^2 \tag{S.4}$$

and $\sigma_{\mathrm{beat}}$ is the expected fluctuation of the beat frequency between the signal and comb waves given as

$$\sigma_{\mathrm{beat}}^2 = \frac{1}{n}\sum_{i=1}^{n}\delta f_i^2 \tag{S.5}$$

The value of $\sigma_{\mathrm{meas}}$ dominates that of $\sigma_{\mathrm{beat}}$ except for the very weak transitions. Note that $\sigma$ is not the expected uncertainty of $\bar{\nu}$. The uncertainty in Table S 3 is similarly obtained. The expected error of the combination differences in Table S 3 is given as

$$\sigma_{\mathrm{CD}} = \sqrt{\sigma_{\mathrm{forbidden}}^2 + \sigma_{\mathrm{allowed}}^2} \tag{S.6}$$

where $\sigma_{\mathrm{forbidden}}$ and $\sigma_{\mathrm{allowed}}$ are the uncertainties of the forbidden and allowed transitions.

## II. OODR SPECTROSCOPY RESULTS

The results of the OODR Λ-type measurements are listed in Table S 1. The first column shows the number of the measurement series. The second column shows the rotational quantum number, symmetry and counting number from the HITRAN database for the upper and lower levels of the $2\nu_3 \rightarrow \nu_3$ transition, and the transition frequency and uncertainty determined in this work. The third and fouth columns show the center frequencies with uncertainties of the $2\nu_3$ transition from Refs. [3,7] and the $\nu_3$ transition from Refs. [8,9], used to calculate the ground state combination difference. The fifth column shows the rotational quantum number and symmetry of the ground state term values involved in the combination difference, the value of the ground state combination difference $\delta\nu_{GS}$ determined in our work with its uncertainty, calculated as the root mean square of the uncertainties of the three involved transitions, see Eq. S.2, and the difference between the measured combination differences and those calculated using the Hamiltonian fit results, excluding measurement series 14 and 15.

Table S 1. List of measured Λ-type OODR transitions from this work, and the reference values used to calculate the combination differences of the ground state terms.

| Series | $2\nu_3 \rightarrow \nu_3$ (this work): State *d* *J* | *C* | HITRAN # | State *c* *J* | *C* | HITRAN # | $\nu_{dc}$ [kHz] | $\sigma_{dc}$ [kHz] | $2\nu_3$: $\nu_{ad}$ [kHz] | $\sigma_{ad}$ [kHz] | $\nu_3$: $\nu_{bc}$ [kHz] | $\sigma_{bc}$ [kHz] | Combination differences: $GS_a$ *J* | *C* | $GS_b$ *J* | *C* | $\delta\nu_{GS}$ [kHz] | $\sigma_{GS}$ [kHz] | Meas. – Calc. [kHz] |
|---|---|---|---|---|---|---|---|---|---|---|---|---|---|---|---|---|---|---|---|
| 1 | 1 | $A_2$ | 18 | 2 | $A_1$ | 4 | 88 858 195 640.1 | 2.1 | [a)]180 345 064 527.5 | 0.23 | [c)]89 601 828 651.6 | 2.5 | 0 | $A_1$ | 3 | $A_2$ | 1 885 040 235.9 | 3.3 | 8.7 |
| 2 | 2 | $F_2$ | 81 | 3 | $F_1$ | 15 | 88 537 047 100.4 | 1.8 | [a)]180 661 736 318.9 | 0.44 | [c)]89 297 695 263.7 | 2.1 | 1 | $F_1$ | 4 | $F_2$ | 2 826 993 954.9 | 2.8 | –2.6 |
| 3 | 3 | E | 71 | 4 | E | 12 | 88 213 485 941.0 | 2.1 | [a)]180 974 330 437.9 | 0.33 | [c)]88 992 600 885.8 | 1.9 | 2 | E | 5 | E | 3 768 243 611.2 | 2.8 | –0.8 |
| 4 | 3 | $F_1$ | 112 | 4 | $F_2$ | 18 | 88 215 663 996.8 | 1.1 | [a)]180 974 442 136.1 | 0.32 | [c)]88 990 490 805.6 | 2.0 | 2 | $F_2$ | 5 | $F_1^{(2)}$ | 3 768 287 333.8 | 2.3 | –2.1 |
| 5 | 4 | $F_2$ | 142 | 5 | $F_1$ | 22 | 87 888 600 281.6 | 2.2 | [b)]181 282 762 617.0 | 8.00 | [c)]88 685 592 346.8 | 2.5 | 3 | $F_1$ | 6 | $F_2^{(2)}$ | 4 708 569 988.7 | 8.7 | 11.6 |
| 6 | 4 | $F_1$ | 136 | 5 | $F_2$ | 21 | 87 892 170 894.1 | 4.8 | [b)]181 283 051 578.0 | 6.00 | [c)]88 682 207 770.5 | 2.4 | 3 | $F_2$ | 6 | $F_1$ | 4 708 672 913.4 | 8.1 | 5.7 |
| 7 | 4 | $A_1$ | 51 | 5 | $A_2$ | 8 | 87 895 528 683.0 | 4.4 | [b)]181 283 407 456.0 | 3.00 | [c)]88 679 126 616.6 | 2.2 | 3 | $A_2$ | 6 | $A_1$ | 4 708 752 156.4 | 5.7 | 9.2 |
| 8 | 5 | $A_2$ | [e)]59 | 6 | $A_1$ | 10 | 87 557 090 481.5 | 1.2 | [a)]181 586 674 001.2 | 0.57 | [c)]88 382 052 721.3 | 2.7 | 4 | $A_1$ | 7 | $A_2$ | 5 647 530 798.4 | 3.0 | 2.8 |
| 9 | 5 | $F_2$ | 167 | 6 | $F_1^{(1)}$ | 25 | 87 548 946 236.2 | 3.7 | [a)]181 587 042 943.7 | 0.43 | [c)]88 391 450 553.3 | 2.1 | 4 | $F_1$ | 7 | $F_2^{(1)}$ | 5 646 646 154.3 | 4.2 | –3.6 |
| 10 | 5 | $F_2$ | 167 | 6 | $F_1^{(2)}$ | 24 | 87 563 075 410.3 | 1.9 | [a)]181 587 042 943.7 | 0.43 | [c)]88 376 181 600.5 | 2.1 | 4 | $F_1$ | 7 | $F_2^{(2)}$ | 5 647 785 932.9 | 2.9 | –2.3 |
| 11 | 5 | E | 113 | 6 | E | 17 | 87 566 242 092.7 | 1.7 | a)181 587 304 103.5 | 1.32 | c)88 373 149 030.1 | 2.4 | 4 | E | 7 | E | 5 647 912 980.7 | 3.2 | –0.6 |
| 12 | 5 | $F_1$ | 175 | 6 | $F_2^{(2)}$ | 26 | 87 571 174 005.7 | 0.7 | a)181 588 087 225.2 | 0.90 | c)88 368 863 360.1 | 2.5 | 4 | $F_2$ | 7 | $F_1^{(2)}$ | 5 648 049 859.4 | 2.8 | 0.9 |
| 13 | 2 | $F_2$ | 81 | 3 | $F_1$ | 13 | 88 652 011 571.4 | 3.6 | a)180 661 736 318.9 | 0.44 | d)91 381 337 559.8 | 2.5 | 1 | $F_1$ | 2 | $F_2$ | 628 387 187.7 | 4.4 | –4.3 |
| 14 | 3 | $F_1$ | 112 | 4 | $F_2$ | 16 | 88 362 732 029.9 | 2.0 | a)180 974 442 136.1 | 0.32 | d)91 669 360 660.9 | 2.9 | 2 | $F_2$ | 3 | $F_1$ | *942 349 445.4 | 3.5 | 14.1 |
| 15 | 4 | $A_1$ | 51 | 5 | $A_2$ | 7 | 88 072 089 294.4 | 1.1 | b)181 283 407 456.0 | 3.00 | d)91 955 347 354.2 | 3.1 | 3 | $A_2$ | 4 | $A_1$ | *1 255 970 807.4 | 4.5 | 20.9 |

a) Ref. [7].
b) Ref. [3].
c) Ref. [9] with the correction of the second-order Doppler shift.
d) Ref. [8] with the correction of the second-order Doppler shift.
e) This state does not have a counting number in HITRAN, therefore we used the value from ExoMol.

# III. LAMB DIP SPECTROSCOPY RESULTS

## 1. Transitions frequencies

Table S 2 shows all allowed and forbidden Lamb-dip transitions measured in this work and in previous works [8-10] included in the Hamiltonian fit. The first and third columns show the rotational quantum numbers and symmetry labels in the ground state, frequencies of the forbidden and allowed transitions, and uncertainties, respectively. The second column lists the quantum numbers of the common upper levels, labeled with the energy order from the HITRAN database [11], where $R$ is the quantum number for end-over-end rotation of the molecule. The fourth column lists the obtained values of the ground state combination differences $\delta\nu_{GS}$ with uncertainties and differences with respect to the Hamiltonian fit. Because the wavefunctions of the rovibrational levels with $J'$ of 12 and 13 are considerably mixed, the forbidden transitions may have absorption intensity comparable with the allowed transitions. Therefore, for some pairs of the allowed and forbidden transitions, the allowed transition is assigned as the stronger one. The Lamb-dip transitions that do not share the upper state with any other transitions are listed in Table S 3.

Table S 2. List of the Lamb-dip frequencies from this and previous works used to calculate the combination differences of the ground state terms.

| Forbidden $\nu_3$ transition | | | | Common upper level | | | | Allowed $\nu_3$ transition | | | | Combination differences | | |
|---|---|---|---|---|---|---|---|---|---|---|---|---|---|---|
| $J$ | $C$ | Frequency [kHz] | $\sigma_{forbidden}$ [kHz] | $J$ | $R$ | C | HITRAN # | $J$ | $C$ | Frequency [kHz] | $\sigma_{allowed}$ [kHz] | Absolute value [kHz] | $\sigma_{CD}$ [kHz] | Meas.–Calc. [kHz] |
| 9 | $F_1^{(3)}$ | 93 336 421 002.1 | 10.9 | 10 | 9 | $F_2$ | 37 | 9 | $F_1^{(2)}$ | 93 337 911 563.1 | 2.4 | 1 490 560.9 | 11.2 | –8.4 |
| 9 | $F_1^{(2)}$ | 93 335 918 226.8 | 10.9 | 10 | 9 | $F_2$ | 36 | 9 | $F_1^{(3)}$ | 93 334 427 653.7 | 4.8 | 1 490 573.1 | 11.9 | 3.8 |
| 10 | $F_1^{(2)}$ | 93 606 120 665.2 | 7.4 | 11 | 10 | $F_2$ | 39 | 10 | $F_1^{(1)}$ | 93 608 383 513.7 | 5.6 | 2 262 848.4 | 9.3 | –3.3 |
| 10 | $F_1^{(1)}$ | 93 606 347 875.9 | 9.9 | 11 | 10 | $F_2$ | 38 | 10 | $F_1^{(2)}$ | 93 604 085 025.6 | 5.6 | 2 262 850.3 | 11.4 | –1.4 |
| 11 | $F_1^{(2)}$ | 93 873 002 562.0 | 11.4 | 12 | 11 | $F_2$ | 43 | 11 | $F_1^{(3)}$ | 93 870 228 207.6 | 3.2 | 2 774 354.4 | 11.8 | –10.9 |
| 11 | $F_1^{(3)}$ | 93 869 530 937.7 | 11.1 | 12 | 11 | $F_2$ | 42 | 11 | $F_1^{(2)}$ | 93 872 305 304.8 | 3.8 | 2 774 367.1 | 11.7 | 1.8 |
| 12 | $F_2^{(2)}$ | 94 137 285 266.5 | 8.6 | 13 | 12 | $F_1$ | 47 | 12 | $F_2^{(3)}$ | 94 134 052 744.8 | 8.2 | 3 232 521.7 | 11.9 | –12.8 |
| 12 | $F_2^{(2)}$ | 94 140 057 589.0 | 9.3 | 13 | 12 | $F_1$ | 48 | 12 | $F_2^{(3)}$ | 94 136 825 063.6 | 3.2 | 3 232 525.4 | 9.8 | –9.1 |
| 10 | $F_2^{(2)}$ | 93 606 390 268.7 | 46.6 | 11 | 10 | $F_1$ | 39 | 10 | $F_2^{(3)}$ | 93 602 932 438.6 | 3.1 | 3 457 830.0 | 46.7 | 0.9 |
| 11 | $F_2^{(3)}$ | 93 872 839 280.7 | 5.5 | 12 | 11 | $F_1$ | 42 | 11 | $F_2^{(2)}$ | 93 877 058 067.6 | 5.2 | 4 218 786.9 | 7.6 | 0.2 |
| 11 | $F_2^{(2)}$ | 93 875 430 964.3 | 6.3 | 12 | 11 | $F_1$ | 41 | 11 | $F_2^{(3)}$ | 93 871 212 167.7 | 2.1 | 42 18 796.6 | 6.6 | 9.9 |
| 12 | $F_2^{(1)}$ | 94 141 145 646.4 | 10.6 | 13 | 12 | $F_1$ | 46 | 12 | $F_2^{(2)}$ | 94 136 901 142.2 | 1.8 | 4 244 504.2 | 10.8 | 4.0 |
| 12 | $F_2^{(2)}$ | 94 137 285 266.5 | 8.6 | 13 | 12 | $F_1$ | 47 | 12 | $F_2^{(1)}$ | 94 141 529 776.0 | 6.8 | 4 244 509.5 | 11.0 | 9.3 |
| 12 | $F_2^{(2)}$ | 94 140 057 589.0 | 9.3 | 13 | 12 | $F_1$ | 48 | 12 | $F_2^{(1)}$ | 94 144 302 104.2 | 8.3 | 4 244 515.1 | 12.5 | 14.9 |
| 11 | $E^{(2)}$ | 93 872 339 849.8 | 8.3 | 12 | 11 | E | 29 | 11 | $E^{(1)}$ | 93 876 940 189.6 | 3.6 | 4 600 339.8 | 9.0 | –9.8 |
| 11 | $E^{(1)}$ | 93 874 727 935.0 | 7.9 | 12 | 11 | E | 28 | 11 | $E^{(2)}$ | 93 870 127 575.4 | 4.2 | 4 600 359.6 | 8.9 | 10.0 |
| 12 | $F_1^{(1)}$ | 94 150 664 409.4 | 31.8 | 13 | 12 | $F_2$ | 45 | 12 | $F_1^{(2)}$ | 94 143 576 595.3 | 6.9 | 7 087 814.1 | 32.5 | –37.8 |
| 12 | $F_1^{(1)}$ | 94 148 519 199.1 | 22.8 | 13 | 12 | $F_2$ | 44 | 12 | $F_1^{(2)}$ | 94 141 431 357.6 | 6.8 | 7 087 841.5 | 23.8 | –10.4 |
| 12 | $F_1^{(3)}$ | 94 134 020 888.8 | 8.1 | 13 | 12 | $F_2$ | 44 | 12 | $F_1^{(2)}$ | 94 141 431 357.6 | 6.8 | 7 410 468.8 | 10.6 | –4.1 |
| 12 | $F_1^{(2)}$ | 94 143 576 595.3 | 6.9 | 13 | 12 | $F_2$ | 45 | 12 | $F_1^{(3)}$ | 94 136 166 112.3 | 5.0 | 7 410 483.0 | 8.5 | 10.1 |
| 12 | $F_2^{(3)}$ | 94 134 052 744.8 | 8.2 | 13 | 12 | $F_1$ | 47 | 12 | $F_2^{(1)}$ | 94 141 529 776.0 | 6.8 | 7 477 031.2 | 10.7 | –3.6 |
| 12 | $F_2^{(1)}$ | 94 144 302 104.2 | 8.3 | 13 | 12 | $F_1$ | 48 | 12 | $F_2^{(3)}$ | 94 136 825 063.6 | 3.2 | 7 477 040.6 | 8.9 | 5.8 |
| 11 | $F_1^{(1)}$ | 93 880 541 323.6 | 24.1 | 12 | 11 | $F_2$ | 43 | 11 | $F_1^{(3)}$ | 93 870 228 207.6 | 3.2 | 10 313 116.0 | 24.3 | –8.5 |
| 11 | $F_1^{(3)}$ | 93 876 800 805.1 | 29.9 | 12 | 11 | $F_2$ | 44 | 11 | $F_1^{(1)}$ | 93 887 113 928.0 | 4.2 | 10 313 122.9 | 30.2 | –1.6 |
| 12 | $F_1^{(1)}$ | 94 150 664 409.4 | 31.8 | 13 | 12 | $F_2$ | 45 | 12 | $F_1^{(3)}$ | 94 136 166 112.3 | 5.0 | 14 498 297.1 | 32.2 | –27.6 |
| 12 | $F_1^{(1)}$ | 94 148 519 199.0 | 22.8 | 13 | 12 | $F_2$ | 44 | 12 | $F_1^{(3)}$ | 94 134 020 888.7 | 8.1 | 14 498 310.3 | 24.2 | –14.4 |
| 12 | $F_1^{(3)}$ | 94 139 118 314.9 | 7.8 | 13 | 12 | $F_2$ | 46 | 12 | $F_1^{(1)}$ | 94 153 616 641.1 | 5.3 | 14 498 326.2 | 9.4 | 1.5 |
| 12 | $A_1^{(2)}$ | 94 138 897 326.5 | 9.2 | 13 | 12 | $A_2$ | 17 | 12 | $A_1^{(1)}$ | 94 153 814 131.4 | 2.0 | 14 916 804.9 | 9.4 | –12.9 |
| 12 | $A_1^{(1)}$ | 94 149 514 077.1 | 9.8 | 13 | 12 | $A_2$ | 16 | 12 | $A_1^{(2)}$ | 94 134 597 252.9 | 4.8 | 14 916 824.2 | 10.9 | 6.5 |
| 3 | $F_2$ | 91 739 747 182.6 | 42.8 | 4 | 4 | $F_1$ | 16 | 4 | $F_2$ | [c] 90 483 521 743.0 | 2.0 | 1 256 225 439.7 | 42.9 | –29.9 |
| 3 | $F_2$ | [b] 90 553 920 731.9 | 12.6 | 3 | 4 | $F_1$ | 15 | 4 | $F_2$ | [c] 89 297 695 263.7 | 2.1 | 1 256 225 468.2 | 12.8 | –1.4 |
| 4 | E | [b] 90 562 349 009.5 | 12.3 | 4 | 5 | E | 12 | 5 | E | [c] 88 992 600 885.8 | 1.9 | 1 569 748 123.7 | 12.4 | –23.9 |
| 4 | E | 92 040 008 290.1 | 25.1 | 5 | 5 | E | 14 | 5 | E | [c] 90 470 260 140.0 | 3.0 | 1 569 748 150.1 | 25.2 | 2.5 |
| 4 | $F_1$ | [b] 90 560 330 445.7 | 14.2 | 4 | 5 | $F_2$ | 18 | 5 | $F_1^{(2)}$ | [c] 88 990 490 805.6 | 1.9 | 1 569 839 640.1 | 14.3 | –35.2 |
| 4 | $F_1$ | 92 041 971 159.3 | 38.2 | 5 | 5 | $F_2$ | 20 | 5 | $F_1^{(2)}$ | [c] 90 472 131 488.7 | 2.0 | 1 569 839 670.6 | 38.3 | –4.7 |
| 5 | $F_2$ | [b] 90 568 553 931.1 | 13.2 | 5 | 6 | $F_1$ | 22 | 6 | $F_2^{(2)}$ | [c] 88 685 592 346.8 | 2.5 | 1 882 961 584.3 | 13.4 | –8.0 |
| 5 | $F_2$ | 92 337 182 259.2 | 16.6 | 6 | 6 | $F_1$ | 23 | 6 | $F_2^{(2)}$ | [c] 90 454 220 661.3 | 2.2 | 1 882 961 597.9 | 16.8 | 5.6 |
| 5 | $F_1^{(1)}$ | [b] 90 565 411 736.1 | 13.4 | 5 | 6 | $F_2$ | 21 | 6 | $F_1$ | [c] 88 682 207 770.5 | 2.4 | 1 883 203 965.6 | 13.6 | –15.6 |
| 5 | $F_1^{(1)}$ | 92 340 957 387.5 | 43.5 | 6 | 6 | $F_2$ | 25 | 6 | $F_1$ | [c] 90 457 753 390.8 | 2.3 | 1 883 203 996.7 | 43.6 | 15.5 |
| 6 | $A_2$ | [b] 90 577 562 991.3 | 12.0 | 6 | 7 | $A_1$ | 10 | 7 | $A_2$ | [c] 88 382 052 721.3 | 3.2 | 2 195 510 270.0 | 12.4 | –4.4 |
| 7 | $A_2$ | [c] 88 238 845 036.2 | 2.5 | 6 | 6 | $A_1$ | 9 | 6 | $A_2$ | [c] 90 434 355 307.8 | 2.0 | 2 195 510 271.6 | 3.2 | –2.8 |
| 6 | $A_2$ | [d] 92 627 835 270.2 | 5.0 | 7 | 7 | $A_1$ | 9 | 7 | $A_2$ | [c] 90 432 324 995.7 | 1.9 | 2 195 510 274.5 | 5.3 | 0.1 |
| 6 | $A_2$ | 92 627 835 278.6 | 5.4 | 7 | 7 | $A_1$ | 9 | 7 | $A_2$ | [c] 90 432 324 995.7 | 1.9 | 2 195 510 283.0 | 5.8 | 8.6 |
| 7 | $A_2$ | [d] 90 321 364 410.2 | 50.0 | 7 | 6 | $A_1$ | 8 | 6 | $A_2$ | [b] 92 516 874 711.8 | 2.7 | 2 195 510 301.6 | 50.1 | 27.2 |
| 6 | $F_2^{(1)}$ | [b] 90 572 146 826.9 | 12.0 | 6 | 7 | $F_1$ | 24 | 7 | $F_2^{(2)}$ | [c] 88 376 181 600.5 | 2.1 | 2 195 965 226.4 | 12.2 | –5.6 |
| 7 | $F_2^{(2)}$ | [c] 88 236 103 827.0 | 5.1 | 6 | 6 | $F_1$ | 22 | 6 | $F_2^{(1)}$ | [c] 90 432 069 056.8 | 3.2 | 2 195 965 229.8 | 6.0 | –2.2 |
| 6 | $F_2^{(1)}$ | 92 632 572 878.8 | 13.2 | 7 | 7 | $F_1$ | 29 | 7 | $F_2^{(2)}$ | [c] 90 436 607 630.2 | 2.1 | 2 195 965 248.7 | 13.3 | 16.7 |
| 6 | E | [b] 90 569 328 656.4 | 11.9 | 6 | 7 | E | 17 | 7 | E | [c] 88 373 149 030.1 | 2.4 | 2 196 179 626.3 | 12.1 | –6.5 |
| 6 | E | 92 635 563 939.8 | 10.6 | 7 | 7 | E | 18 | 7 | E | [c] 90 439 384 289.5 | 2.4 | 2 196 179 650.3 | 10.9 | 17.5 |
| 7 | E | 92 934 249 293.8 | 32.0 | 8 | 8 | E | 22 | 8 | $E^{(2)}$ | [b] 90 426 450 988.0 | 2.5 | 2 507 798 305.9 | 32.1 | –72.7 |
| 7 | $F_2^{(1)}$ | [b] 90 580 089 029.8 | 13.3 | 7 | 8 | $F_1$ | 31 | 8 | $F_2^{(1)}$ | [b] 88 072 045 716.9 | 2.5 | 2 508 043 312.9 | 13.5 | –3.7 |
| 7 | $F_2^{(1)}$ | 92 919 927 072.8 | 11.5 | 8 | 8 | $F_1$ | 30 | 8 | $F_2^{(1)}$ | [b] 90 411 883 753.8 | 3.5 | 2 508 043 319.0 | 12.0 | 2.4 |
| 7 | $F_1^{(1)}$ | [b] 90 572 602 576.6 | 16.4 | 7 | 8 | $F_2$ | 28 | 8 | $F_1^{(2)}$ | [b] 88 063 885 766.1 | 2.8 | 2 508 716 810.5 | 16.6 | –19.1 |
| 7 | $F_1^{(1)}$ | 92 925 796 140.0 | 12.4 | 8 | 8 | $F_2$ | 32 | 8 | $F_1^{(2)}$ | [b] 90 417 079 295.1 | 3.2 | 2 508 716 844.9 | 12.8 | 15.3 |
| 8 | $F_1^{(2)}$ | 93 224 693 991.9 | 39.4 | 9 | 9 | $F_2$ | 35 | 9 | $F_1^{(3)}$ | [b] 90 405 139 447.8 | 2.7 | 2 819 554 544.1 | 39.5 | 6.9 |
| 8 | $E^{(1)}$ | [b] 90 582 066 017.0 | 13.2 | 8 | 9 | E | 23 | 9 | E | [b] 87 762 118 602.0 | 2.4 | 2 819 947 415.0 | 13.4 | 0.7 |
| 8 | $E^{(1)}$ | 93 207 417 582.7 | 14.6 | 9 | 9 | E | 23 | 9 | E | [b] 90 387 470 162.7 | 3.3 | 2 819 947 420.0 | 15.0 | 5.7 |
| 8 | $F_1^{(1)}$ | [b] 90 580 618 339.4 | 11.8 | 8 | 9 | $F_2$ | 34 | 9 | $F_1^{(2)}$ | [b] 87 760 424 092.0 | 2.1 | 2 820 194 247.4 | 12.0 | –7.6 |
| 8 | $F_1^{(1)}$ | 93 209 135 618.2 | 8.7 | 9 | 9 | $F_2$ | 34 | 9 | $F_1^{(2)}$ | [b] 90 388 941 358.9 | 3.1 | 2 820 194 259.3 | 9.2 | 4.3 |
| 8 | $A_1$ | [b] 90 572 599 170.0 | 13.0 | 8 | 9 | $A_2$ | 14 | 9 | $A_1$ | [b] 87 751 536 553.2 | 2.2 | 2 821 062 616.8 | 13.2 | –14.6 |
| 8 | $A_1$ | 93 214 580 512.9 | 8.5 | 9 | 9 | $A_2$ | 13 | 9 | $A_1$ | [b] 90 393 517 876.1 | 2.4 | 2 821 062 636.8 | 8.9 | 5.3 |
| 9 | $A_1$ | [d] 90 252 069 280.2 | 30.0 | 9 | 8 | $A_2$ | 12 | 8 | $A_1$ | [b] 93 073 131 926.1 | 3.9 | 2 821 062 645.9 | 30.3 | 14.4 |
| 9 | $A_1$ | [b] 90 571 670 600.1 | 11.8 | 9 | 10 | $A_2$ | 14 | 10 | $A_1$ | [b] 87 442 078 690.9 | 2.8 | 3 129 591 909.1 | 12.1 | 0.4 |
| 9 | $A_1$ | 93 506 277 314.7 | 13.5 | 10 | 10 | $A_2$ | 13 | 10 | $A_1$ | [b] 90 376 685 394.0 | 2.7 | 3 129 591 920.7 | 13.7 | 12.0 |
| 9 | $F_2^{(1)}$ | [b] 90 582 766 396.7 | 12.8 | 9 | 10 | $F_1$ | 38 | 10 | $F_2^{(2)}$ | [b] 87 451 442 423.4 | 2.4 | 3 131 323 973.2 | 13.0 | –1.5 |
| 9 | $F_2^{(1)}$ | 93 491 633 700.6 | 10.5 | 10 | 10 | $F_1$ | 37 | 10 | $F_2^{(2)}$ | [b] 90 360 309 721.1 | 3.7 | 3 131 323 979.6 | 11.1 | 4.9 |
| 9 | $F_1^{(1)}$ | [b] 90 580 299 279.0 | 11.8 | 9 | 10 | $F_2$ | 37 | 10 | $F_1^{(1)}$ | [b] 87 448 447 306.5 | 2.5 | 3 131 851 972.4 | 12.1 | –25.7 |
| 9 | $F_1^{(1)}$ | 93 494 448 830.5 | 14.3 | 10 | 10 | $F_2$ | 39 | 10 | $F_1^{(1)}$ | [b] 90 362 596 828.1 | 3.5 | 3 131 852 002.4 | 14.7 | 4.3 |
| 10 | $F_1^{(1)}$ | 93 792 440 523.0 | 20.6 | 11 | 11 | $F_2$ | 41 | 11 | $F_1^{(2)}$ | [b] 90 351 646 408.9 | 2.9 | 3 440 794 114.2 | 20.9 | –28.4 |
| 10 | $A_2$ | 93 771 627 374.1 | 12.4 | 11 | 11 | $A_1$ | 13 | 11 | $A_2$ | [b] 90 329 810 743.6 | 3.8 | 3 441 816 630.5 | 12.9 | 4.1 |
| 10 | $A_2$ | [b] 90 583 538 390.0 | 12.1 | 10 | 11 | $A_1$ | 15 | 11 | $A_2$ | [b] 87 141 721 757.0 | 2.5 | 3 441 816 632.9 | 12.4 | 6.5 |
| 10 | $F_2^{(1)}$ | 93 773 621 589.4 | 18.8 | 11 | 11 | $F_1$ | 43 | 11 | $F_2^{(3)}$ | [b] 90 331 277 255.9 | 3.5 | 3 442 344 333.5 | 19.1 | –8.2 |
| 10 | $E^{(1)}$ | 93 774 798 122.1 | 10.4 | 11 | 11 | E | 28 | 11 | $E^{(1)}$ | [b] 90 332 130 944.2 | 2.7 | 3 442 667 177.9 | 10.7 | 3.4 |
| 10 | $F_2^{(2)}$ | 93 802 305 477.3 | 32.0 | 11 | 11 | $F_1$ | 44 | 11 | $F_2^{(3)}$ | [b] 90 359 262 570.0 | 3.1 | 3 443 042 907.3 | 32.2 | –79.3 |
| 4 | $A_1$ | 92 131 907 977.8 | 25.3 | 5 | 6 | $A_2$ | 8 | 6 | $A_1$ | [c] 88 679 126 616.6 | 2.2 | 3 452 781 361.2 | 25.4 | 0.4 |
| 11 | $F_2^{(1)}$ | 94 050 537 851.8 | 11.5 | 12 | 12 | $F_1$ | 44 | 12 | $F_2^{(1)}$ | [b] 90 298 367 659.6 | 2.8 | 3 752 170 192.3 | 11.8 | –0.9 |
| 11 | $F_1^{(1)}$ | 94 052 146 039.6 | 14.0 | 12 | 12 | $F_2$ | 46 | 12 | $F_1^{(2)}$ | [b] 90 299 370 976.3 | 2.8 | 3 752 775 063.3 | 14.2 | –0.8 |
| 11 | $A_2$ | 94 086 685 882.5 | 31.9 | 12 | 12 | $A_1$ | 17 | 12 | $A_2$ | [b] 90 332 479 746.8 | 3.4 | 3 754 206 135.7 | 32.1 | –0.4 |
| 11 | $A_2$ | [b] 90 551 117 470.7 | 13.7 | 11 | 12 | $A_1$ | 14 | 12 | $A_2$ | [b] 86 796 911 324.1 | 3.4 | 3 754 206 146.5 | 14.1 | 10.4 |
| 5 | $F_1^{(1)}$ | 92 447 669 782.8 | 26.7 | 6 | 7 | $F_2$ | 26 | 7 | $F_1^{(2)}$ | [b] 88 368 863 360.1 | 2.5 | 4 078 806 422.7 | 26.8 | 21.2 |
| 7 | $A_2$ | 93 068 138 503.9 | 20.8 | 8 | 9 | $A_1$ | 12 | 9 | $A_2$ | [b] 87 740 265 613.2 | 2.9 | 5 327 872 890.8 | 21.0 | 27.5 |
| 8 | $A_1$ | 93 392 733 204.0 | 39.9 | 9 | 10 | $A_2$ | 14 | 10 | $A_1$ | [b] 87 442 078 690.9 | 2.8 | 5 950 654 513.0 | 40.0 | –27.1 |
| 8 | $F_2^{(1)}$ | 93 376 391 489.5 | 34.9 | 9 | 10 | $F_1$ | 37 | 10 | $F_2^{(3)}$ | [b] 87 425 525 323.4 | 2.2 | 5 950 866 166.0 | 35.0 | 29.4 |
| 10 | $A_1$ | 93 982 445 449.8 | 30.4 | 11 | 12 | $A_2$ | 15 | 12 | $A_1^{(2)}$ | [b] 86 789 799 611.9 | 2.2 | 7 192 645 837.8 | 30.5 | –3.9 |

[a] The intense transition between a pair of transitions is called allowed, and the other is forbidden.
[b] Ref. [8] with the correction of the second-order Doppler shift.
[c] Ref. [9] with the correction of the second-order Doppler shift.
[d] Ref. [10] with the correction of the second-order Doppler shift.

Table S 3. The 17 Lamb-dip transitions that do not share the upper level with other observed Lamb-dip transitions.

| Lower level | | Upper level | | | | | |
|---|---|---|---|---|---|---|---|
| *J* | *C* | *J* | *R* | *C* | HITRAN # | Frequency [kHz] | Uncertainty [kHz] |
| 9 | $A_2$ | 10 | 9 | $A_1$ | 13 | 93 333 228 710.6 | 2.0 |
| 9 | $F_2^{(2)}$ | 10 | 9 | $F_1$ | 34 | 93 333 781 970.1 | 6.5 |
| 9 | $A_1$ | 10 | 9 | $A_2$ | 12 | 93 336 790 815.9 | 4.0 |
| 9 | E | 10 | 9 | E | 24 | 93 338 231 629.1 | 3.0 |
| 9 | $F_2^{(1)}$ | 10 | 9 | $F_1$ | 35 | 93 346 601 982.1 | 2.8 |
| 9 | $F_1^{(1)}$ | 10 | 9 | $F_2$ | 38 | 93 346 738 518.1 | 4.9 |
| 10 | $E^{(2)}$ | 11 | 10 | E | 26 | 93 603 751 151.6 | 6.0 |
| 10 | $A_1$ | 11 | 10 | $A_2$ | 14 | 93 605 157 975.3 | 2.6 |
| 10 | $F_2^{(2)}$ | 11 | 10 | $F_1$ | 40 | 93 608 521 472.2 | 3.7 |
| 10 | $A_2$ | 11 | 10 | $A_1$ | 12 | 93 618 090 905.3 | 5.5 |
| 10 | $F_2^{(1)}$ | 11 | 10 | $F_1$ | 41 | 93 618 121 375.5 | 5.2 |
| 10 | $E^{(1)}$ | 11 | 10 | E | 27 | 93 618 132 283.4 | 4.7 |
| 11 | $A_2$ | 12 | 11 | $A_1$ | 16 | 93 876 484 045.3 | 3.3 |
| 11 | $F_2^{(1)}$ | 12 | 11 | $F_1$ | 43 | 93 887 092 195.9 | 3.5 |
| 12 | $E^{(2)}$ | 13 | 12 | E | 30 | 94 137 340 438.2 | 3.1 |
| 12 | $A_2$ | 13 | 12 | $A_1$ | 15 | 94 137 529 018.1 | 1.5 |
| 12 | $E^{(1)}$ | 13 | 12 | E | 31 | 94 153 507 128.2 | 4.8 |

## IV. MICROWAVE AND RF DATA USED IN THE HAMILTONIAN FIT

### RF and MW spectroscopy in the ground state

Table S 4 lists the RF and MW transitions in the ground state from the previous works [12-14] with residuals of the weighted least-square fit. The first and second columns list the rotational quantum number and the symmetry label of the lower and upper levels.

Table S 4 RF and MW transitions in the ground state

| Lower level | | Upper level | | | | |
|---|---|---|---|---|---|---|
| *J* | *C* | *J* | *C* | Frequency [kHz] | Uncertainty [kHz] | Meas.–Calc. [kHz] |
| 2 | E | 2 | $F_2$ | [a] 7 970.30 | 0.66 | –0.2 |
| 7 | $F_2^{(2)}$ | 7 | $F_1^{(2)}$ | [b] 423 020 | 20 | –8.1 |
| 7 | $F_1^{(1)}$ | 7 | $F_2^{(2)}$ | [b] 1 246 550 | 20 | 3.5 |
| 11 | $F_1^{(1)}$ | 11 | $F_2^{(3)}$ | [c] 9 361 072 | 50 | 43.4 |
| 12 | $F_1^{(1)}$ | 12 | $F_2^{(2)}$ | [c] 10 797 985 | 50 | 119.5 |
| 12 | $F_1^{(1)}$ | 12 | $F_2^{(3)}$ | [c] 14 030 436 | 50 | 35.9 |
| 12 | $A_1^{(1)}$ | 12 | $A_2$ | [c] 13 279 651 | 50 | 9.6 |
| 12 | $E^{(1)}$ | 12 | $E^{(2)}$ | [c] 10 321 942 | 50 | 23.8 |
| 12 | $F_2^{(1)}$ | 12 | $F_1^{(3)}$ | [c] 7 944 957 | 50 | –2.4 |

[a] Ref. [13].
[b] Ref. [12].
[c] Ref. [14].

**MW spectroscopy in the $v_3 = 1$ state**

Table S 5 shows the combination differences determined from MW transitions in the $v_3 = 1$ state [15,16] and the $\nu_3$ band transitions [8,9]. The first column lists the rotational quantum number, the angular momentum quantum number for end-over-end rotation of the molecule, and the symmetry label of the upper and lower levels of the MW transition in the $v_3 = 1$ state. The second and third columns list the rotational quantum number and the symmetry label in the lower level of the allowed $\nu_3$ band transitions. The fourth column lists the determined CD with the expected error and the residual of the fit.

Table S 5 MW transitions in the $v_3 = 1$ state and the derived combination differences in the ground state

| MW transition in $v_3$= 1 | | | | | | | | IR from GS to lower level in $v_3$=1 | | | | IR from GS to upper level in $v_3$=1 | | | GS Combination Difference | | |
|---|---|---|---|---|---|---|---|---|---|---|---|---|---|---|---|---|---|
| | | Lower level | | Upper level | | Frequency | Uncertainty | | | Frequency | Uncertainty | | Frequency | Uncertainty | [f] Det. | Expected error | Meas.–Calc. |
| *J* | *R* | *C* | HITRAN# | *C* | HITRAN# | [kHz] | [kHz] | *J* | *C* | [kHz] | [kHz] | *C* | [kHz] | [kHz] | [kHz] | [kHz] | [kHz] |
| 4 | 4 | $F_2$ | 17 | $F_1$ | 16 | [a] 11 867 042 | [b] 13 | 4 | $F_1$ | [c] 90 471 813 803.6 | 1.9 | $F_2$ | [c] 90 483 521 743.0 | 2.0 | 159 102.6 | 13.3 | –2.3 |
| 5 | 5 | $F_1$ | 21 | $F_2$ | 20 | [a] 17 080 899 | [b] 5 | 5 | $F_2$ | [c] 90 455 380 115.4 | 2.5 | $F_1^{(2)}$ | [c] 90 472 131 488.7 | 2.0 | 329 525.7 | 5.9 | 5.9 |
| 8 | 8 | $F_2$ | 32 | $F_1$ | 31 | [a] 12 035 064 | [b] 5 | 8 | $F_1^{(2)}$ | [d] 90 417 079 295.1 | 3.2 | $F_2^{(2)}$ | [d] 90 428 470 084.9 | 3.0 | 644 274.2 | 6.7 | –2.3 |
| 12 | 12 | $A_1$ | 17 | $A_2$ | 15 | [a] 14 393 265 | [b] 5 | 12 | $A_2$ | [d] 90 332 479 746.8 | 3.4 | $A_1^{(2)}$ | [d] 90 345 235 846.7 | 2.9 | 1 637 165.1 | 6.7 | –11.2 |
| 5 | 6 | $A_2$ | 8 | $A_1$ | 6 | [a] 14 801 356 | [b] 6 | 6 | $A_1$ | [c] 88 679 126 616.6 | 2.2 | $A_2$ | [c] 88 694 688 813.9 | 3.2 | 760 841.3 | 7.1 | 1.7 |
| 5 | 6 | $F_2$ | 21 | $F_1$ | 23 | [a] 15 126 853 | [b] 4 | 6 | $F_1$ | [c] 88 682 207 770.5 | 2.4 | $F_2^{(1)}$ | [c] 88 698 120 463.0 | 2.0 | 785 839.5 | 5.1 | –0.2 |
| 6 | 7 | $F_1$ | 24 | $F_2$ | 27 | [a] 15 601 844 | [b] 4 | 7 | $F_2^{(2)}$ | [c] 88 376 181 600.5 | 2.1 | $F_1^{(1)}$ | [c] 88 393 029 988.6 | 2.2 | 1 246 544.1 | 5.0 | –2.4 |
| 7 | 8 | $F_1$ | 31 | $F_2$ | 29 | [a] 11 872 725 | [b] 7 | 8 | $F_2^{(1)}$ | [d] 88 072 045 716.9 | 2.5 | $F_1^{(1)}$ | [d] 88 085 481 994.7 | 2.4 | 1 563 552.8 | 7.8 | 9.6 |
| 8 | 9 | $F_2$ | 34 | $F_1$ | 33 | [a] 14 226 562 | [b] 4 | 9 | $F_1^{(2)}$ | [d] 87 760 424 092.0 | 2.1 | $F_2^{(1)}$ | [d] 87 777 197 025.8 | 2.4 | 2 546 371.8 | 5.1 | –7.2 |
| 8 | 9 | $F_1$ | 32 | $F_2$ | 34 | [a] 15 885 433 | [b] 8 | 9 | $F_2^{(2)}$ | [d] 87 742 746 571.2 | 2.9 | $F_1^{(2)}$ | [d] 87 760 424 092.0 | 2.1 | 1 792 087.8 | 8.8 | –3.3 |
| 9 | 10 | $F_2$ | 37 | $F_1$ | 39 | [a] 16 629 252 | [b] 8 | 10 | $F_1^{(1)}$ | [d] 87 448 447 306.5 | 2.5 | $F_2^{(1)}$ | [d] 87 469 023 275.6 | 3.4 | 3 946 717.1 | 9.0 | 0.7 |
| 9 | 10 | $F_2$ | 36 | $F_1$ | 38 | [a] 17 390 793 | [b] 10 | 10 | $F_1^{(2)}$ | [d] 87 431 362 196.1 | 2.6 | $F_2^{(2)}$ | [d] 87 451 442 423.4 | 2.4 | 2 689 434.3 | 10.6 | 8.1 |
| 10 | 11 | $F_1$ | 39 | $F_2$ | 42 | [a] 12 229 867 | [b] 11 | 11 | $F_2^{(3)}$ | [d] 87 114 104 427.1 | 3.7 | $F_1^{(2)}$ | [d] 87 128 156 558.3 | 2.6 | 1 822 264.2 | 11.9 | –5.2 |
| 10 | 11 | $F_1$ | 40 | $F_2$ | 43 | [a] 15 497 275 | [b] 9 | 11 | $F_2^{(2)}$ | [d] 87 139 103 696.5 | 2.7 | $F_1^{(1)}$ | [d] 87 159 743 216.3 | 2.7 | 5 142 244.8 | 9.8 | 2.9 |
| 11 | 12 | $F_2$ | 43 | $F_1$ | 46 | [a] 17 334 963 | [b] 13 | 12 | $F_1^{(3)}$ | [d] 86 791 594 163.8 | 2.6 | $F_2^{(2)}$ | [d] 86 812 629 581.7 | 2.4 | 3 700 454.9 | 13.5 | –4.3 |
| 6 | 7 | $F_2$ | 26 | $F_1$ | 24 | [e] 6 895 204 | 10 | 7 | $F_1^{(2)}$ | [c] 88 368 863 360.1 | 2.5 | $F_2^{(2)}$ | [c] 88 376 181 600.5 | 2.1 | 423 036.4 | 10.5 | 8.3 |

[a] Ref. [16].
[b] The paper [a] reported three times the standard deviations.
[c] Ref. [9] with the correction of the second-order Doppler shift.
[d] Ref. [8] with the correction of the second-order Doppler shift.
[e] Ref. [15].
[f] The absolute value of the combination difference determined from MW transitions in the $\upsilon_3 = 1$ state and the allowed $\nu_3$ band transitions.

## V. GROUND STATE TERM VALUES FROM THE HAMILTONIAN FIT

Table S 6 lists the rotational levels of the ground vibrational state of methane obtained from the Hamiltonian fit, together with the expected error.

Table S 6 Ground state term values and their expected errors calculated from the Hamiltonian fit.

| *J* | C | Term value [kHz] | Expected error [kHz] |
|---|---|---|---|
| 1 | $F_1$ | 314 231 927.62 | 0.32 |
| 2 | E | 942 611 149.13 | 0.85 |
| 2 | $F_2$ | 942 619 119.64 | 0.85 |
| 3 | $F_1$ | 1 884 968 551.0 | 1.4 |
| 3 | $F_2$ | 1 885 000 415.5 | 1.4 |
| 3 | $A_2$ | 1 885 040 227.2 | 1.4 |
| 4 | $A_1$ | 3 141 011 013.6 | 2.0 |
| 4 | $F_1$ | 3 141 066 780.2 | 1.9 |
| 4 | E | 3 141 106 613.5 | 1.9 |
| 4 | $F_2$ | 3 141 225 885.1 | 1.8 |
| 5 | $F_1^{(1)}$ | 4 710 469 342.1 | 2.4 |
| 5 | $F_2$ | 4 710 576 935.7 | 2.2 |
| 5 | E | 4 710 854 761.1 | 2.0 |
| 5 | $F_1^{(2)}$ | 4 710 906 455.6 | 2.0 |
| 6 | E | 6 592 839 962.0 | 2.9 |
| 6 | $F_2^{(1)}$ | 6 592 887 483.5 | 2.8 |
| 6 | $A_2$ | 6 593 031 534.8 | 2.5 |
| 6 | $F_2^{(2)}$ | 6 593 538 528.0 | 2.1 |
| 6 | $F_1$ | 6 593 673 323.2 | 2.0 |
| 6 | $A_1$ | 6 593 792 374.4 | 2.1 |
| 7 | $F_1^{(1)}$ | 8 787 606 169.0 | 3.4 |
| 7 | $F_2^{(1)}$ | 8 787 712 938.1 | 3.2 |
| 7 | $A_2$ | 8 788 541 809.2 | 2.4 |
| 7 | $F_2^{(2)}$ | 8 788 852 715.5 | 2.2 |
| 7 | E | 8 789 019 594.8 | 2.1 |
| 7 | $F_1^{(2)}$ | 8 789 275 743.6 | 2.1 |
| 8 | $A_1$ | 11 294 094 294.2 | 4.5 |
| 8 | $F_1^{(1)}$ | 11 294 192 711.5 | 4.4 |
| 8 | $E^{(1)}$ | 11 294 250 477.0 | 4.4 |
| 8 | $F_2^{(1)}$ | 11 295 756 254.8 | 3.2 |
| 8 | $F_1^{(2)}$ | 11 296 323 004.9 | 3.1 |
| 8 | $E^{(2)}$ | 11 296 817 973.4 | 2.8 |
| 8 | $F_2^{(2)}$ | 11 296 967 275.1 | 2.8 |
| 9 | $F_1^{(1)}$ | 14 111 739 138.7 | 6.9 |
| 9 | $F_2^{(1)}$ | 14 111 840 587.5 | 7.0 |
| 9 | E | 14 114 197 891.2 | 5.3 |
| 9 | $F_1^{(2)}$ | 14 114 386 966.6 | 5.3 |
| 9 | $A_1$ | 14 115 156 925.6 | 5.8 |
| 9 | $F_1^{(3)}$ | 14 115 877 535.8 | 5.2 |
| 9 | $F_2^{(2)}$ | 14 116 179 057.6 | 5.1 |
| 9 | $A_2$ | 14 116 414 672.4 | 5.1 |
| 10 | $E^{(1)}$ | 17 239 602 740.5 | 10.8 |
| 10 | $F_2^{(1)}$ | 17 239 644 420.4 | 10.9 |
| 10 | $A_2$ | 17 239 734 737.3 | 11.0 |
| 10 | $F_2^{(2)}$ | 17 243 164 562.3 | 8.9 |
| 10 | $F_1^{(1)}$ | 17 243 591 143.8 | 8.9 |
| 10 | $A_1$ | 17 244 748 834.3 | 8.4 |
| 10 | $F_1^{(2)}$ | 17 245 853 988.5 | 9.0 |
| 10 | $E^{(2)}$ | 17 246 217 640.1 | 9.0 |
| 10 | $F_2^{(3)}$ | 17 246 622 391.3 | 8.8 |
| 11 | $F_1^{(1)}$ | 20 676 846 520.3 | 14.6 |
| 11 | $F_2^{(1)}$ | 20 676 916 904.6 | 14.6 |
| 11 | $A_2$ | 20 681 551 363.6 | 12.5 |
| 11 | $F_2^{(2)}$ | 20 681 988 762.6 | 12.7 |
| 11 | $E^{(1)}$ | 20 682 269 915.0 | 12.8 |
| 11 | $F_1^{(2)}$ | 20 684 385 279.4 | 12.1 |
| 11 | $F_2^{(3)}$ | 20 686 207 548.9 | 12.8 |
| 11 | $E^{(2)}$ | 20 686 870 264.6 | 12.5 |
| 11 | $F_1^{(3)}$ | 20 687 159 644.7 | 12.5 |
| 12 | $A_1^{(1)}$ | 24 422 477 858.3 | 16.4 |
| 12 | $F_1^{(1)}$ | 24 422 533 732.5 | 16.3 |
| 12 | $E^{(1)}$ | 24 422 562 573.3 | 16.3 |
| 12 | $F_2^{(1)}$ | 24 429 087 097.8 | 15.4 |
| 12 | $F_1^{(2)}$ | 24 429 621 584.4 | 15.2 |
| 12 | $E^{(2)}$ | 24 432 884 491.5 | 15.7 |
| 12 | $F_2^{(2)}$ | 24 433 331 598.0 | 15.6 |
| 12 | $A_2$ | 24 435 757 499.7 | 15.4 |
| 12 | $F_2^{(3)}$ | 24 436 564 132.6 | 15.4 |
| 12 | $F_1^{(3)}$ | 24 437 032 057.2 | 15.5 |
| 12 | $A_1^{(2)}$ | 24 437 394 676.0 | 15.6 |

Figure S 2 shows the residuals between the ground state energies from the Hamiltonian fit and from a) ExoMol [17] and b) HITRAN2024 [11], where the HITRAN2024 values were retrieved from the data in the region of the $\nu_4$ band. Panel a) shows the same data as Fig. 5 in the text but the vertical scale range is larger to show the error bars for levels with $J > 8$. HITRAN provides the ground-state term values rounded to the nearest $10^{-4}$ cm$^{-1}$ (3 MHz).

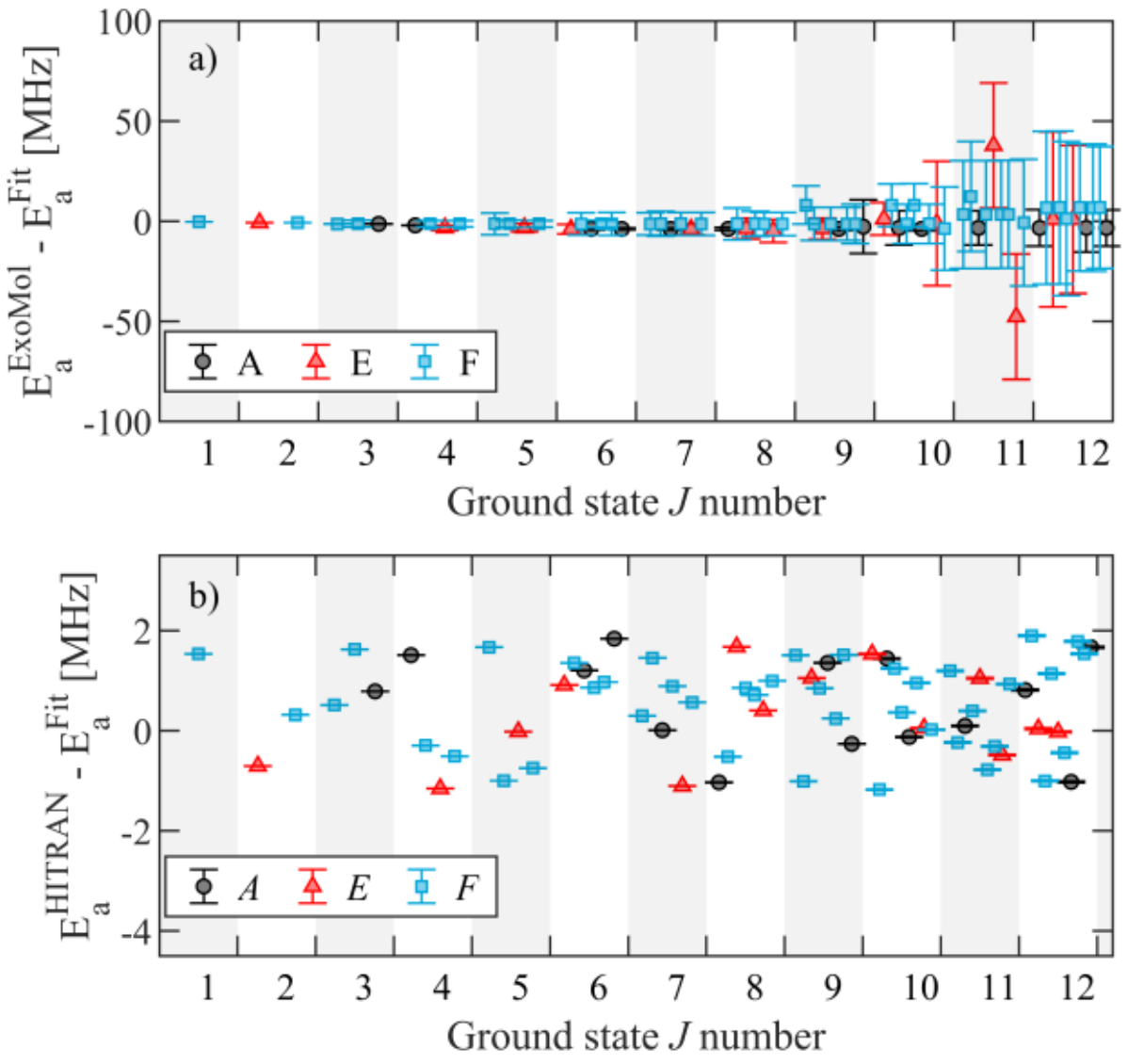

Fig. S2 (Color online) The difference between ground state energies from the Hamiltonian fit and a) ExoMol and b) HITRAN, ordered by the rotational $J$ number and coded for the different symmetries as indicated in the legend.

# REFERENCES

[1] R. W. P. Drever, J. L. Hall, F. V. Kowalski, J. Hough, G. M. Ford, A. J. Munley, and H. Ward, Appl. Phys. B **31**, 97 (1983).
[2] G. Sobon, T. Martynkien, P. Mergo, L. Rutkowski, and A. Foltynowicz, Opt. Lett. **42**, 1748 (2017).
[3] V. Silva de Oliveira, A. Hjältén, I. Silander, A. Rosina, M. Rey, K. K. Lehmann, and A. Foltynowicz, Opt. Express **33**, 38776 (2025).
[4] K. Kefala, V. Boudon, S. N. Yurchenko, and J. Tennyson, J. Quant. Spectr. Rad. Transf. **316**, 108897 (2024).
[5] J. Tennyson *et al.*, J. Quant. Spectr. Rad. Transf. **326**, 109083 (2024).
[6] S. N. Bagaev, E. V. Baklanov, and V. P. Chebotaev, JETP Letters **16**, 344 (1972).
[7] O. Votava, S. Kassi, A. Campargue, and D. Romanini, Phys. Chem. Chem. Phys. **24**, 4157 (2022).
[8] M. Abe, K. Iwakuni, S. Okubo, and H. Sasada, J. Opt. Soc. Am. B **30**, 1027 (2013).
[9] S. Okubo, H. Nakayama, K. Iwakuni, H. Inaba, and H. Sasada, Opt. Express **19**, 23878 (2011).
[10] S. Okubo, H. Inaba, S. Okuda, and H. Sasada, Phys. Rev. A **103**, 022809 (2021).
[11] I. E. Gordon *et al.*, J. Quant. Spectr. Rad. Transf. **353**, 109807 (2026).
[12] R. F. Curl, J. Mol. Spectrosc. **48**, 165 (1973).
[13] W. M. Itano and I. Ozier, J. Chem. Phys. **72**, 3700 (1980).
[14] M. Oldani, M. Andrist, A. Bauder, and A. G. Robiette, J. Mol. Spectrosc. **110**, 93 (1985).
[15] M. Takami, K. Uehara, and K. Shimoda, Japan. J. Appl. Phys. **12**, 924 (1973).
[16] C. J. Pursell and D. P. Weliky, J. Mol. Spectrosc. **153**, 303 (1992).
[17] S. N. Yurchenko, A. Owens, K. Kefala, and J. Tennyson, Mon. Not. Royal Astron. Soc. **528**, 3719 (2024).